\documentclass[british]{scrartcl}
\usepackage[T1]{fontenc}
\usepackage{textcomp}
\usepackage[utf8]{inputenc}
\usepackage{geometry}
\usepackage{babel}
\usepackage{cprotect}
\usepackage{float}
\usepackage{amsmath}
\usepackage{amsthm}
\usepackage{amssymb}
\usepackage{graphicx}
\usepackage{setspace}
\usepackage{esint}
\usepackage[numbers]{natbib}
\usepackage[unicode=true,pdfusetitle,
 bookmarks=true,bookmarksnumbered=false,bookmarksopen=false,
 breaklinks=false,pdfborder={0 0 1},backref=false,colorlinks=false]
 {hyperref}

\makeatletter

\providecommand{\tabularnewline}{\\}

\newcommand{\lyxaddress}[1]{
	\par {\raggedright #1
	\vspace{1.4em}
	\noindent\par}
}

\makeatother

\begin{document}
\title{A Polynomial Approach to the Spectrum of Dirac-Weyl Polygonal Billiards}
\author{M. F. C. Martins Quintela\thanks{mfcmquintela@fc.up.pt}\: and J.
M. B. Lopes dos Santos}
\maketitle

\lyxaddress{Centro de Física das Universidades do Minho e do Porto, CF-UM-UP
e Departmento de Física e Astronomia, Universidade do Porto, Rua do
Campo Alegre 4169-007, Porto, Portugal.}
\begin{abstract}
The Schr{\"o}dinger equation in a square or rectangle with hard walls
is solved in every introductory quantum mechanics course. Solutions
for other polygonal enclosures only exist in a very restricted class
of polygons, and are all based on a result obtained by Lamé in 1852.
Any enclosure can, of course, be addressed by finite element methods
for partial differential equations. In this paper, we present a variational
method to approximate the low-energy spectrum and wave-functions for
arbitrary convex polygonal enclosures, developed initially for the
study of vibrational modes of plates. In view of the recent interest
in the spectrum of quantum dots of two dimensional materials, described
by effective models with massless electrons, we extend the method
to the Dirac-Weyl equation for a spin-1/2 fermion confined in a quantum
billiard of polygonal shape, with different types of boundary conditions.
We illustrate the method's convergence in cases where the spectrum
in known exactly and apply it to cases where no exact solution exists.
\end{abstract}
\global\long\def\hati{\hat{\mbox{\textbf{ı}}}}%
\global\long\def\hatj{\hat{\mbox{\textbf{\j}}}}%
\global\long\def\hatk{\hat{\mbox{\textbf{k}}}}%
\global\long\def\ket#1{\left|#1\right\rangle }%
\global\long\def\bra#1{\left\langle #1\right|}%
\global\long\def\braket#1#2{\langle#1|#2\rangle}%
\newcommand{\squeezeup}{\vspace{-2.5mm}}

\section{Introduction}

One of the first problems solved in introductory quantum mechanics
courses is that of particle enclosed in a square or rectangular box
with hard walls---Dirichlet boundary conditions, $\psi\left(\mathbf{r}\right)=0$
at the boundary. Separation of variables easily turns the problem
into two one-dimensional problems and the solutions are a finite sum
of plane waves. The circle and the ellipse are also amenable to an
exact treatment, again with separation of variables. But the vast
majority of students are never led to consider the case of general
polygonal enclosures. 

The eigenvalue problem of the Schr{\"o}dinger equation for an equilateral
triangle with hard boundaries was solved, long before Schr{\"o}dinger's
equation even came into existence, by Lamé {[}\citealp{Lame1852}{]}
in 1852. This solution has been revisited often from different perspectives
{[}\citealp{ISI:A1980KL96700005,ISI:000183800600005,ISI:A1991GC36900022,li_particle_1987,gaddah_lie_2013}{]}.
As in the case of a rectangle, Lamé's solution is also a finite sum
of plane waves, built to satisfy the boundary conditions, and remarkably,
it has even been proven that the only shapes that share this feature
are the square, the rectangle, the equilateral triangle, the half-square
(right isosceles triangle) and the half equilateral triangle (30-60º
triangle) {[}\citealp{ISI:A1993LQ34500002}{]}.

Following the pioneering work of Berry and Mondragon {[}\citealp{berry_michael_victor_neutrino_1987}{]},
and the discovery that two-dimensional (2D) materials, such as graphene,
can be described by effective continuum models with the Dirac-Weyl
equation {[}\citealp{novoselov_two-dimensional_2005,castro_neto_electronic_2009}{]},
there has been a upsurge of interest in the properties of confined
relativistic particles (Dirac Billiards), either because low energy
solutions have possible relevance for quantum dots of 2D materials
{[}\citealp{ponomarenko_chaotic_2008,libisch_graphene_2009,zarenia_energy_2011,gaddah_exact_2018}{]},
or to address fundamental questions concerning the classical-quantum
relations and energy-level statistics in the semi-classical limit
{[}\citealp{ISI:000289149100002,ISI:000521256300001,ISI:A1995RG85100019,berry_michael_victor_neutrino_1987}{]}.

Lacking exact analytical solutions, one usually has to resort to numerical
methods, and finite element methods of integration of partial differential
equations have universal application {[}\citealp{zarenia_energy_2011}{]}.
But such methods give little physical insight into the solutions.
The aim of this article is to present an essentially variational method
that can yield the low energy spectrum of convex polygonal enclosures,
both in the Schr{\"o}dinger and Dirac equation cases, with the added
benefit of providing approximate wave-functions which are polynomials.
The method is well known in the Mechanical Engineering literature,
in the context of the Helmholtz equation {[}\citealp{bhat_flexural_1987,liew_free_1990,liew_set_1991}{]},
where it was developed for the study of vibration of polygonal plates.
The extension to the Schr{\"o}dinger equation for free particle is
trivial---the eigenvalue problem is the same as the Helmholtz problem---,
but, to our knowledge, its extension to the Dirac-Weyl equation has
not been presented before. 

This paper is structured as follows. In section \ref{sec:2}, we present
the polynomial method for the simple case of the Schr{\"o}dinger
equation, more specifically for the textbook problem of the particle
confined to a square well, and assess the convergence of the spectrum
against the exact solution for this problem. We then apply the method
to an hexagonal enclosure, a problem that has no analytical solution.

Section \ref{sec:3} reviews the different types of boundary conditions
(BC) for the Dirac-Weyl equation, following closely reference {[}\citealp{berry_michael_victor_neutrino_1987}{]}.
We extend the polynomial method to 2-component spinors and the different
types of BC, and illustrate it in exactly solvable one-dimensional
enclosures. In sections \ref{sec:5}--\ref{subsec:Dirichlet-Hexagonal-Enclosure}
we study three different 2D polygonal enclosures with this method,
the circle, the square and the hexagon. For a specific boundary condition,
the solution of the former can be obtained from the solution of Lamé
{[}\citealp{Lame1852}{]}, and we compare our numerical results with
the analytical ones {[}\citealp{gaddah_exact_2018}{]}.

\section{The Polynomial Method for convex polygonal enclosures\protect\label{sec:2}}

\subsection{The method: general idea}

The essential idea of the polynomial method {[}\citealp{bhat_flexural_1987,liew_free_1990,liew_set_1991}{]}
is to generate orthonormal polynomials of increasing order that satisfy,
from the start, the required boundary conditions (BC). In the $Oxy$
plane a linear polynomial in $x$ and $y$, $\mathcal{P}_{1}\left(x,y\right)=a_{0}+a_{1}x+a_{2}y$
will be zero in a straight line, $a_{0}+a_{1}x+a_{2}y=0$. It follows
that a product of $n$ such polynomials trivially satisfies the condition
$\mathcal{P}_{n}\left(x,y\right)=0$ at the boundary of a polygonal
convex region. After normalization, such a function is a zero-th order
approximation to the ground state. 

As an example, for a right triangle with vertexes $\left(0,0\right),\,\left(1,0\right),\,\left(0,1\right)$
we would choose 
\begin{align}
\psi_{0}\left(x,y\right) & =N_{0}xy\left(1-x-y\right)
\end{align}
 with
\begin{align*}
N_{0} & =\left[\int_{0}^{1}dx\int_{0}^{1-x}dy\left[xy\left(1-x-y\right)\right]^{2}\right]^{-1/2}\\
 & =12\sqrt{35}
\end{align*}
If we multiply $\psi_{0}\left(x,y\right)$ by any polynomial, $\psi\left(x,y\right)=\mathcal{P}\left(x,y\right)\psi_{0}\left(x,y\right)$,
we end up with a function that still satisfies the BC. We therefore
proceed to generate a set of linearly independent functions 
\begin{equation}
\psi_{n}\left(x,y\right)=\mathcal{P}_{n}\left(x,y\right)\psi_{0}\left(x,y\right)
\end{equation}
where $\left\{ \mathcal{P}_{n}\left(x,y\right):n=0,1,\dots\right\} $
is a family of ascending-order monomials/polynomials. These can be
generated by taking into account the symmetries of the enclosure in
question, as a simple ordering of monomials, or as a generalization
of Legendre polynomials discussed by Larcher {[}\citealp{larcher_notes_1959}{]}.
In {[}\citealp{liew_set_1991}{]} the authors resort to the sorting
\begin{equation}
\left\{ 1,x,y,xy,x^{2},y^{2},x^{2}y,y^{2}x,x^{2}y^{2},x^{3},...\right\} .\label{eq:liew_monom}
\end{equation}
We truncate the series at a finite $n$ and orthonormalize the functions,
typically through Gram-Schmidt orthogonalization. The Hamiltonian
matrix in this truncated basis is calculated and diagonalized. For
regular polygons, one can take into account the symmetry group of
the enclosure by replacing the monomials of Eq.(\ref{eq:liew_monom})
by increasing order polynomials that transform according to irreducible
representations of the symmetry group {[}\citealp{Quintela2019}{]}.
The Hamiltonian matrix becomes block diagonal, and the resulting eigenstates
end up classified according to their symmetry properties. 

The problem of the Schr{\"o}dinger equation for a particle confined
to a square well, provides a suitable example to assess convergence. 

\subsection{The square well: convergence}

The exact spectrum of the Schr{\"o}dinger equation for a particle
confined to a square well is 
\begin{equation}
E_{n_{1},n_{2}}=\frac{\hbar^{2}}{2m}\frac{\pi^{2}}{L^{2}}\left(n_{1}^{2}+n_{2}^{2}\right),\label{eq:exact_eigvals}
\end{equation}
where $L$ is the side length of the enclosure, and $n_{1(2)}$ positive
integers.

\begin{figure}[H]
\noindent \centering{}\includegraphics[scale=0.33]{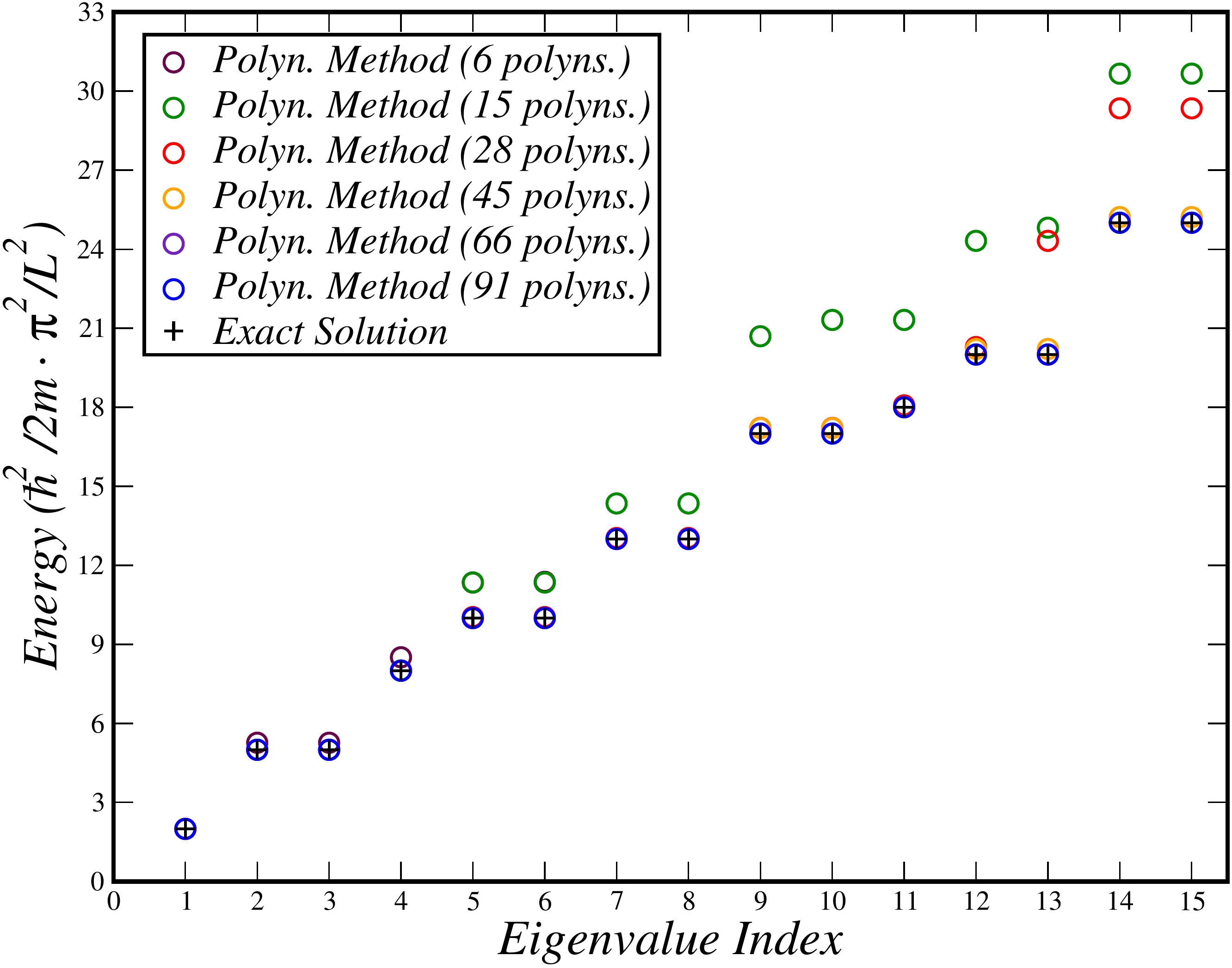}\caption{Convergence of the eigenvalues of the Hamiltonian in a square enclosure
(polynomial method versus exact solution).\protect\label{fig:Convergence-of-orthogonal_square-2}}
\end{figure}

Centering the enclosure at the origin, the starting wavefunction is
\begin{equation}
\psi_{0}\left(x,y\right)=N_{0}\left[\left(\frac{L}{2}\right)^{2}-x^{2}\right]\left[\left(\frac{L}{2}\right)^{2}-y^{2}\right].\label{eq:square_poly}
\end{equation}
The higher order basis functions are generated as described previously,
by multiplication by $x^{m}y^{n}$ monomials. Both the orthogonalization
and the calculation of the matrix elements of the kinetic energy operator
reduce to the calculation of integrals of $x^{m}y^{n}$ monomials
in the domain of the enclosure; these are easily calculated symbolically
and stored. Keeping the computation symbolic up until the diagonalization
of the Hamiltonian avoids stability problems of the Gram-Schmidt procedure.
We computed the Hamiltonian matrix for different basis sizes, and
diagonalized it numerically using standard packages. The results are
shown in Fig.\ref{fig:Convergence-of-orthogonal_square-2}, overlayed
with the exact eigenvalues.

The method is seen to converge towards the exact eigenvalues as we
increase the basis size. As the index of the eigenvalue increases,
we require a larger basis, as expected. The numerical results indicate,
roughly, that to the get the lowest $n$ eigenvalues to within 1\%
accuracy requires of the order of $3n$ basis functions.

\subsection{The hexagonal enclosure}

To apply this procedure to the Schr{\"o}dinger equation for a particle
confined to an hexagonal well of side length $L$, we define the initial
normalized polynomial as 
\begin{align}
\psi_{0}\left(x,y\right) & =N_{0}\left(\frac{3L^{2}}{4}-y^{2}\right)\left[L^{2}-\left(x-\frac{y}{\sqrt{3}}\right)^{2}\right]\nonumber \\
 & \ \ \ \times\left[L^{2}-\left(x+\frac{y}{\sqrt{3}}\right)^{2}\right].
\end{align}

\begin{figure}[H]
\noindent \centering{}\includegraphics[scale=0.33]{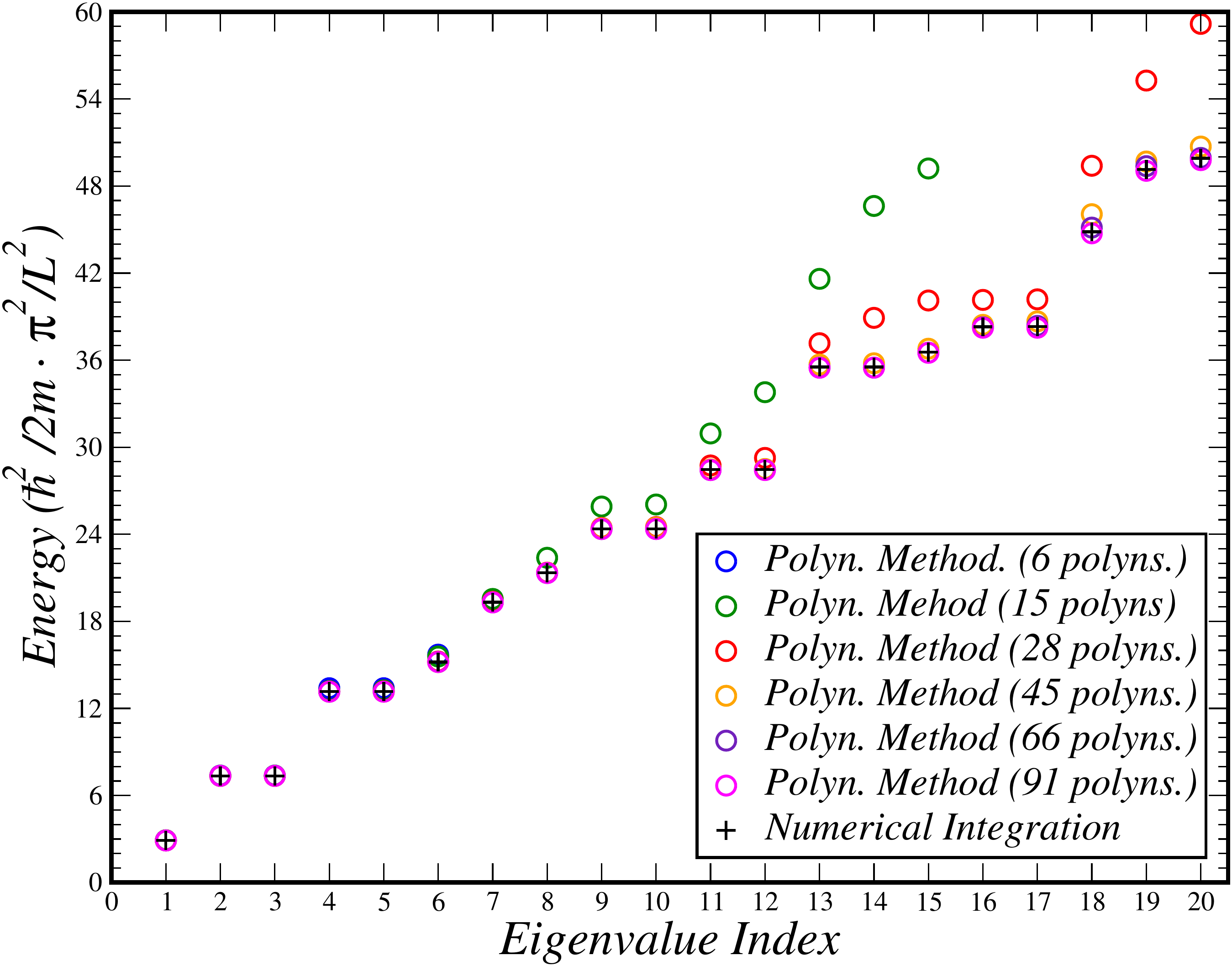}\caption{Convergence of the eigenvalues of the Hamiltonian in an hexagonal
enclosure (polynomial method versus direct numerical integration of
PDE).\protect\label{fig:Convergence-of-orthogonal_square-1}}
\end{figure}

We again calculated the Hamiltonian matrix elements for different
basis sizes, and diagonalized it numerically using standard packages.
The results are shown in Fig.\ref{fig:Convergence-of-orthogonal_square-1},
overlayed with the eigenvalues obtained via direct numerical integration
of the partial differential equation (PDE), by finite elements method.
The convergence is similar to the one found in the case of the square. 

\begin{figure}[H]
\noindent \centering{}\includegraphics[scale=0.33]{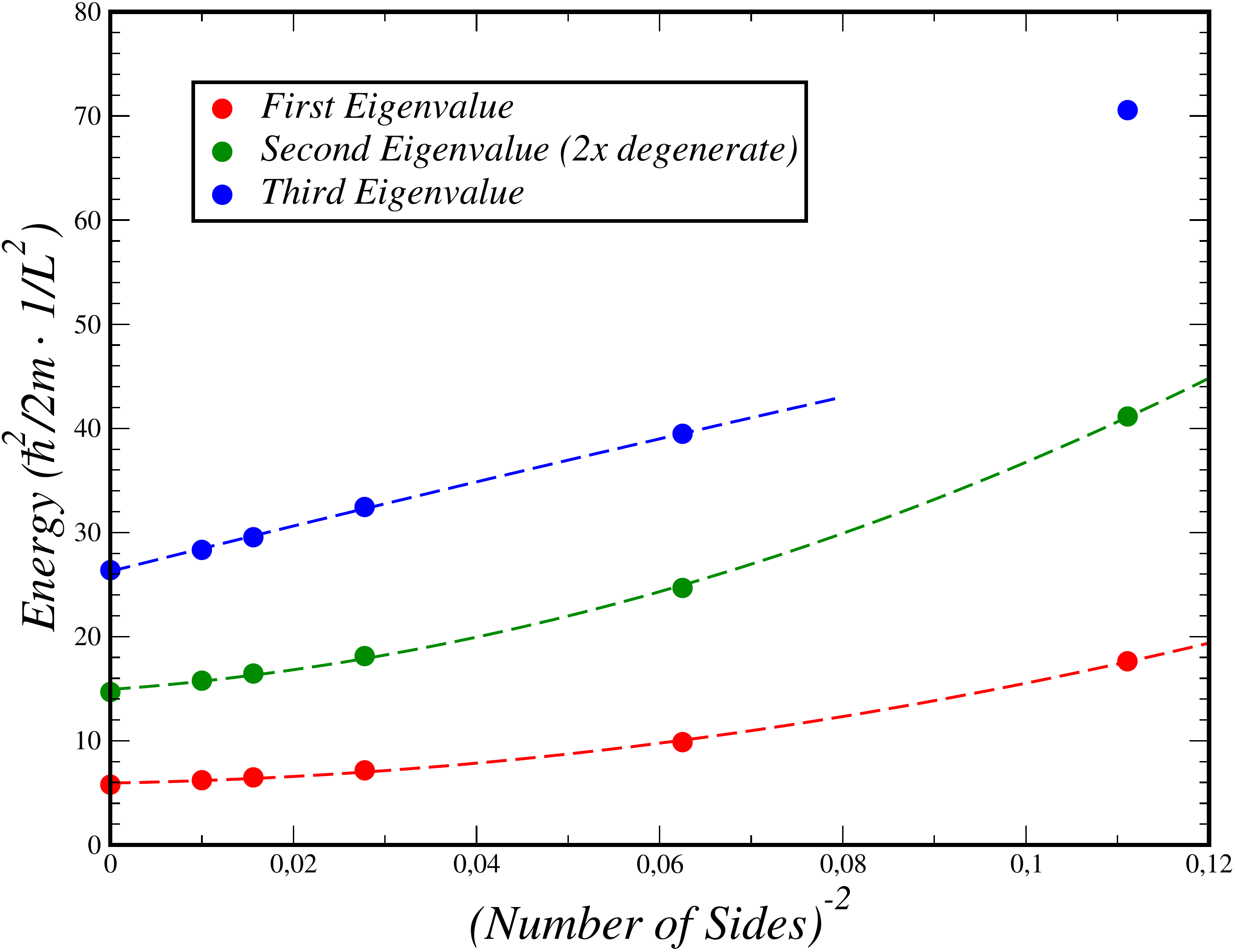}\caption{\protect\label{fig:Number_of_sides-1}Comparison of the first three
distinct energy levels as a function of the inverse of the number
of sides of the polygon squared.}
\end{figure}

As an illustration of this method, we applied it to different polygons
with increasing number of sides and plot the lowest eigenvalues against
the inverse number of sides (squared) in Fig.\ref{fig:Number_of_sides-1},
where the result for the circle, $\epsilon_{i}^{\infty},i=1,2,3$,
is the analytical solution. 
The dashed lines are fits of the form $\epsilon_{i}(n)=\epsilon_{i}^{\infty}+a_{2}n^{-2}+a_{4}n^{-4}$,
where $n$ is the number of sides. A suitable fit could not be found
for the third eigenvalue, and the point for $n=3$ was excluded. The
remaining data points are well fitted with $a_{4}=0.$ 

We now turn to the extension of these results to the Dirac-Weyl equation. 

\section{Confinement of Dirac--Weyl electrons\protect\label{sec:3}}

\subsection{Boundary Conditions in Neutrino Billiards}

Berry and Mondragon {[}\citealp{berry_michael_victor_neutrino_1987}{]}
were the first to address the issue of the boundary conditions of
a two-component spinor solution of the Dirac-Weyl equation in a confining
planar enclosure. The eigenvalue equation is 
\begin{equation}
-i\hbar v_{F}\boldsymbol{\mathbf{\sigma}}\cdot\mathbf{\nabla}\Psi=E\Psi.
\end{equation}
where $\boldsymbol{\mathbf{\sigma}}=\left(\sigma_{x},\sigma_{y}\right)$.
We assume a confining boundary at $y=0$, excluding the region $y<0$.
The probability current is given by 
\begin{equation}
\mathbf{j}=v_{F}\Psi^{\dagger}\boldsymbol{\mathbf{\sigma}}\Psi=v_{F}\left\langle \boldsymbol{\mathbf{\sigma}}\right\rangle ,
\end{equation}
which means that a necessary condition for confined states is 
\begin{equation}
\left\langle \sigma^{y}\right\rangle _{\mathrm{boundary}}=0.
\end{equation}
This condition implies that the spinor at the boundary must be an
eigenstate of $\sigma_{u}=\mathbf{\sigma}\cdot\mathbf{u}$, where
$\mathbf{u}=\cos\theta\mathbf{e_{z}}+\sin\theta\mathbf{e_{x}}$. \emph{i.e}.,
\begin{equation}
\left[\begin{array}{cc}
\cos\theta & \sin\theta\\
\sin\theta & -\cos\theta
\end{array}\right]\left[\begin{array}{c}
\psi_{A}\\
\psi_{B}
\end{array}\right]=\eta\left[\begin{array}{c}
\psi_{A}\\
\psi_{B}
\end{array}\right],\ \eta=\pm1,
\end{equation}
or simply 
\begin{align}
\frac{\psi_{B}}{\psi_{A}} & =\frac{\eta-\cos\theta}{\sin\theta}.
\end{align}

Denoting $t:=\tan\left(\theta/2\right)$, simple trigonometric identities
lead us to
\begin{equation}
\frac{\psi_{B}}{\psi_{A}}=t,\label{eq:_t_BC}
\end{equation}
 with a spinor, at the boundary, given by 
\begin{equation}
\left[\begin{array}{c}
\psi_{A}\\
\psi_{B}
\end{array}\right]=\frac{1}{\sqrt{1+t^{2}}}\left[\begin{array}{c}
1\\
t
\end{array}\right]\label{eq:spinor}
\end{equation}
for $\eta=+1$. The case for $\eta=-1$ is obtained by the substitution
$t\rightarrow-1/t$, or by changing $\theta\rightarrow\pi-\theta$.
As such, this is not a different solution, and we keep only $\eta=+1$. 

Some special cases are:
\begin{itemize}
\item $t=0$ or $t\rightarrow\infty$ (Dirichlet boundaries)
\begin{equation}
\left\{ \begin{array}{ll}
{t=0,} & {\psi_{B}=0};\\
{t\rightarrow\infty,} & {\psi_{A}=0}
\end{array}\right.;
\end{equation}
so that, at the boundary, $\left\langle \sigma^{z}\right\rangle =\pm1$.
\item $t=\pm1$
\begin{equation}
\psi_{A}=\pm\psi_{B}.\label{eq:inf_mass}
\end{equation}
In this case, the spin points along the $x$-direction.
\end{itemize}
The latter case corresponds to the infinite-mass confinement by Berry
and Mondragon {[}\citealp{berry_michael_victor_neutrino_1987}{]}.
They considered adding a spatially dependent mass (gap) term, 
\begin{equation}
-i\hbar v_{F}\boldsymbol{\mathbf{\sigma}}\cdot\mathbf{\nabla}\Psi+m\left(\mathbf{r}\right)\sigma_{z}=E\Psi.
\end{equation}
and showed that continuity of the spinor components at the boundary,
with $m(\mathbf{r})=0$ inside the enclosure, and $m(\mathbf{r})\to\infty$
outside, leads to BC of the type of Eq.(\ref{eq:inf_mass}). Similarly,
one can show that a mass term of the form {[}\citealp{Quintela2019}{]}
\begin{equation}
m\left(\mathbf{r}\right)\left(\sigma^{z}-\cos\theta\right)
\end{equation}
leads to the more general BC of Eq.(\ref{eq:_t_BC}) with $t=\tan\left(\theta/2\right)$.
The cases $t=0$ or $t\to\pm\infty$ correspond to Dirichlet boundaries
in one of the spinor components, as in the case of zigzag boundaries
in graphene. 

If the boundary is at an angle $\phi$ with the $x$-axis, one must
rotate the spinor in Eq. \ref{eq:spinor}, resulting finally in 
\begin{equation}
\Psi=\frac{e^{-i\phi\sigma_{z}/2}}{\sqrt{1+t^{2}}}\left[\begin{array}{l}
{1}\\
{t}
\end{array}\right]=\frac{1}{\sqrt{1+t^{2}}}\left[\begin{array}{c}
{e^{-i\phi/2}}\\
{e^{i\phi/2}t}
\end{array}\right].\label{eq:alpha_BC}
\end{equation}

We will now show how to extend the polynomial method defined in Section
\ref{sec:2} to this new type of BC. As an illustration we begin with
1D cases, where one easily computes the exact spectrum.

\subsection{One-Dimensional Dirac-Weyl confinement\protect\label{subsec:One-dimensional-Dirac-weyl}}

Consider the simple case where $t=1$ and the confined region is the
interval $y\in\left[-L/2,L/2\right]$. The BC are $\psi_{B}\left(-L/2\right)/\psi_{A}\left(-L/2\right)=1$
at one end, and $\psi_{B}\left(L/2\right)/\psi_{A}\left(L/2\right)=-1,$
at the other ($\phi=\pi$ in Eq.(\ref{eq:alpha_BC})). Measuring energies
in units of $\hbar v_{F}/L$, $\epsilon=E/\left(\hbar v_{F}/L\right)$,
\begin{equation}
\begin{aligned}-\partial_{y}\psi_{B} & =\epsilon\psi_{A}\\
\partial_{y}\psi_{A} & =\epsilon\psi_{B}
\end{aligned}
.
\end{equation}
Squaring the Hamiltonian, one sees that the solutions have the form
\begin{equation}
\begin{aligned}\psi_{B} & =f_{1}e^{iqy}+f_{2}e^{-iqy}\\
\psi_{A} & =f_{3}e^{iqy}+f_{4}e^{-iqy}
\end{aligned}
,
\end{equation}
with $\epsilon=sq$, $s=\pm1$. With no loss of generality we assume
$q>0$. The Dirac-Weyl equation implies
\begin{align*}
f_{4} & =isf_{2}.\\
f_{3} & =-isf_{1}
\end{align*}
The BC then imply
\begin{equation}
\begin{array}{r}
{\left(1-ist\right)f_{1}+\left(1+ist\right)f_{2}e^{-iqL}=0}\\
{\left(1+ist\right)f_{1}+\left(1-ist\right)f_{2}e^{iqL}=0}
\end{array}.
\end{equation}
A non-zero solution to this homogeneous system of 2 equations, for
$t=1$, requires
\begin{equation}
\left(1-is\right)^{2}e^{iqL}=\left(1+is\right)^{2}e^{-iqL}
\end{equation}
or 
\begin{equation}
q=\frac{\pi}{2L}\left(2n+1\right)
\end{equation}
and the energies are
\begin{equation}
\epsilon_{s}(n)=s\hbar v_{F}\frac{\pi}{2L}\left(2n+1\right)\qquad n\geq0,s=\pm1\label{eq:energy_1D}
\end{equation}
The Dirac-Weyl equation in free space has electron-hole symmetry;
it is easily seen that for any solution $\Psi=\left[\psi_{A},\psi_{B}\right]$$^{T}$
of energy $E$, the state $\overline{\Psi}=\left[\psi_{A},-\psi_{B}\right]^{T}$
is a solution of energy $-E$. But the BC we are choosing, $\psi_{B}/\psi_{A}=\pm1$,
breaks this symmetry, because $\Psi$ and $\overline{\Psi}$ cannot
both obey these BC. This can be traced to the mass term we introduced
outside the confining region to derive the BC: it explicitly breaks
this particle-hole symmetry. Nevertheless, the spectrum of eigenvalues
remains symmetrical (Eq.\ref{eq:energy_1D}), because it is still
true for this complete 1D Hamiltonian, inside and outside the enclosure,
that $\sigma_{x}$ is a chiral symmetry operator, $\sigma_{x}\hat{H}\sigma_{x}=-\hat{H}$.

\begin{figure}[H]
\centering{}\includegraphics[scale=0.33]{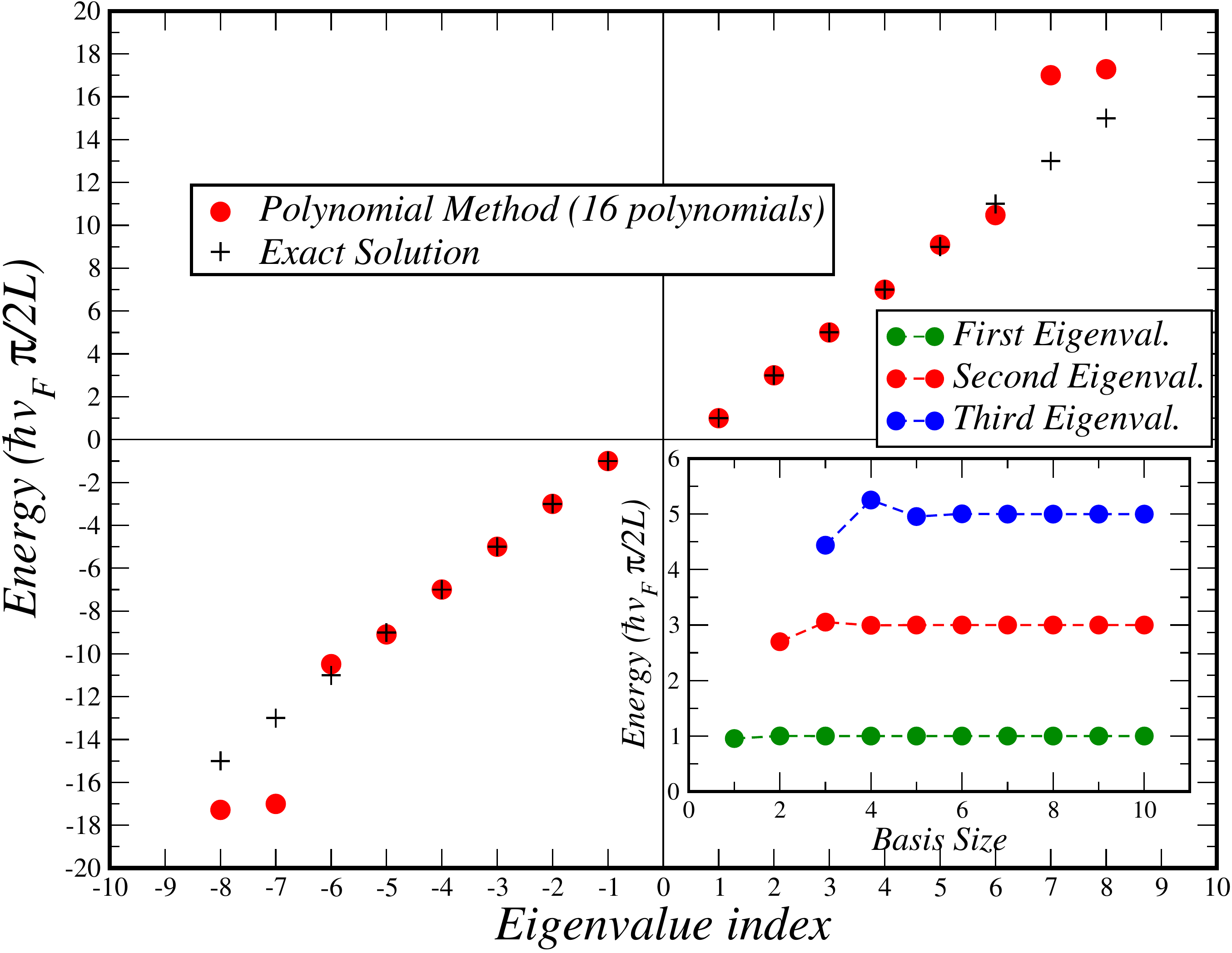}\caption{\protect\label{fig:(Left)-Comparison-between1d-1}Exact solution vs
polynomial approximation for a 1-D strip with $t=1$. Convergence
of the first three eigenvalues in the inset.}
\end{figure}

To implement the polynomial method, we must first realize that the
parities of $\psi_{A}$ and $\psi_{B}$ are opposite. We therefore
choose for the spinor components as the lowest order polynomials that
satisfy the BC. For the conduction band $\left(s=1\right)$ we have,
\begin{equation}
\Psi_{0}^{(1)}=N_{0}\left[\begin{array}{c}
1\\
-\frac{y}{L/2}
\end{array}\right]
\end{equation}
where $N_{0}$ is a normalization constant. To obtain the valence
band solution, $s=-1$, it is enough to swap $\psi_{A}\leftrightarrow\psi_{B}$.
The basis is now generated by functions of the form $\Psi_{n}^{(s)}=\mathcal{P}_{n}(y)\Psi_{0}^{(s)}$,
orthonormalized by the Gram-Schmidt process. One must take care, however,
to orthogonalize states between the valence and conduction bands,
as the states generated this way for $s=1$ and $s=-1$ are not orthogonal
to start with. 

\begin{figure}[H]
\centering{}\includegraphics[scale=0.33]{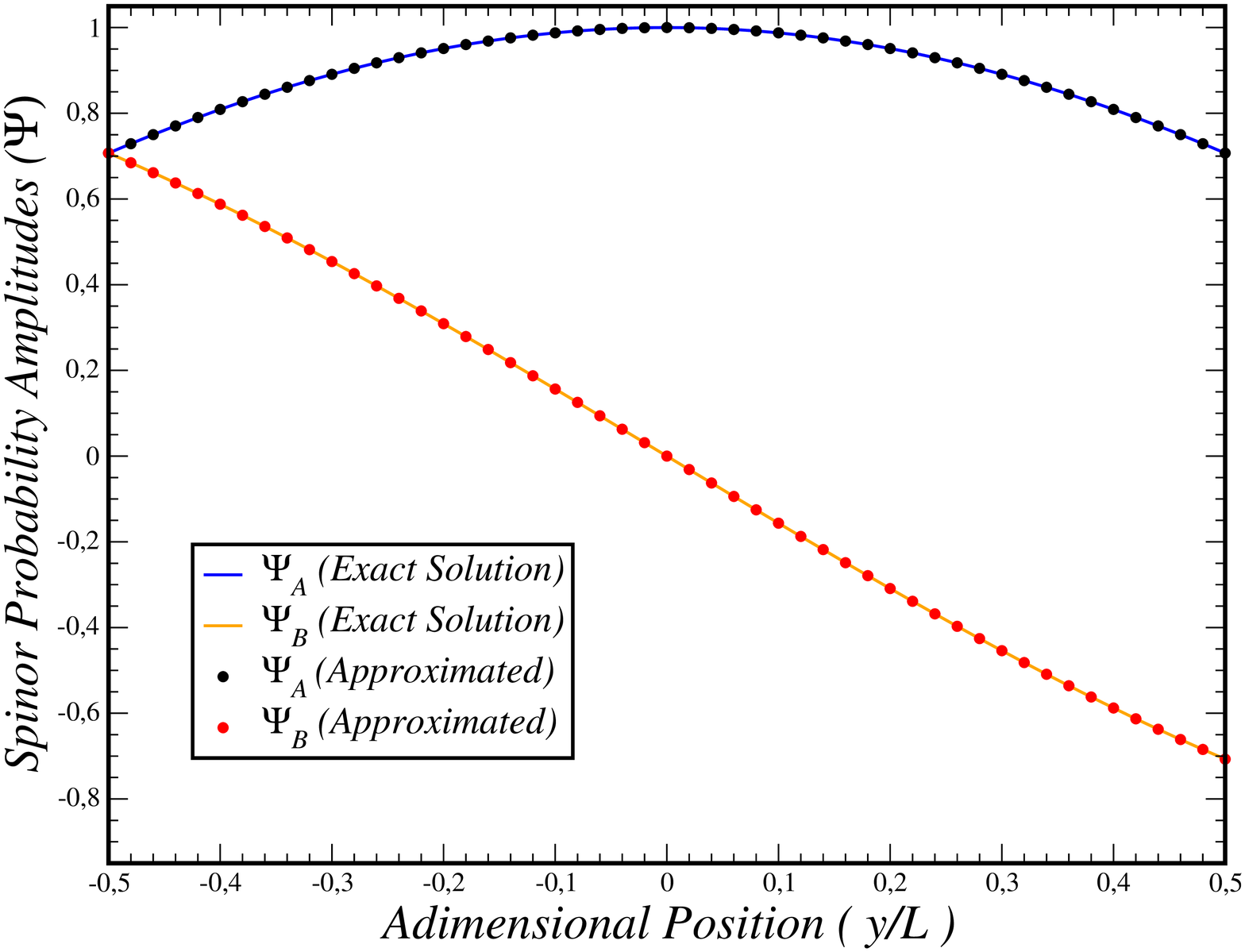}\caption{\protect\label{fig:Comparison-between-the_dirac_lowest}Comparison
between the exact solution (full lines) and the lowest-energy polynomial
eigenfunction (dots) for $16$ polynomials.}
\end{figure}

Having generated the basis, one easily computes and diagonalizes the
Hamiltonian matrix. We obtained the approximate spectrum visible in
Fig.\ref{fig:(Left)-Comparison-between1d-1}; the inset shows the
convergence analysis of the first three eigenvalue of the conduction
band. An almost exact match is also present in the first eigenfunction
of the conduction band, as shown in Fig.\ref{fig:Comparison-between-the_dirac_lowest}. 

Having checked the validity of this method against a simple solution
of the Dirac-Weyl equation, we will now proceed to consider planar
enclosures.

\subsection{Two-Dimensional Enclosures\protect\label{sec:5}}

To generalize this procedure to two-dimensional enclosures, we must
be aware of a difference between Dirichlet boundaries and boundaries
with finite $t$. If one has a Dirichlet BC, one can square the Dirac-Weyl
Hamiltonian and easily prove that each spinor component obeys a Schr{\"o}dinger
equation inside the enclosure where the mass term is zero. As such,
one can use the method we presented before for this equation to calculate
the square of the eigenvalues and the spinor component that is zero
at the boundary. Since the Dirac-Weyl equation has particle-hole symmetry---if
one changes the sign of one of the spinor components in a state with
energy $E$, one obtains a state of energy $-E$, which still preserves
the Dirichlet BC---, one effectively obtains all the eigenvalues.
Applying the Dirac operator to the component which is zero at the
Boundary, one generates the other component of the spinor and completes
the solution {[}\citealp{gaddah_exact_2018}{]}. 

\subsubsection{Non-Dirichlet Boundary Conditions}

The same procedure cannot be used for finite $t$, when the BC only
fixes the \emph{ratio }between the spinor components, $\psi_{B}/\psi_{A}=te^{i\phi}$.
The circle is the only 2D enclosure where an exact solution is known
for the infinite-mass BC we are using {[}\citealp{Quintela2019}{]}.
The solution is given by {[}\citealp{PhysRevB.77.035316}{]}
\begin{equation}
\Psi_{\epsilon,m}\left(r,\theta\right)=\left[\begin{array}{c}
e^{im\theta}\mathcal{J}_{m}\left(q_{\epsilon}r\right)\\
ise^{i\left(m+1\right)\theta}\mathcal{J}_{m+1}\left(q_{\epsilon}r\right)
\end{array}\right]\label{eq:solu_circle}
\end{equation}
where $s=\pm1$ is the band index, $\mathcal{J}_{m}$ is the Bessel
function of the first kind of order $m$ and $\epsilon=s\hbar v_{F}q_{\epsilon}$
is the energy of the state. Imposing the boundary condition, the allowed
energy levels are those in which 
\begin{equation}
\frac{\mathcal{J}_{m+1}\left(q_{\epsilon}R\right)}{\mathcal{J}_{m}\left(q_{\epsilon}R\right)}=s.
\end{equation}

To apply the polynomial method to this enclosure, we must build the
initial functions carefully. The rotation symmetry is continuous:
\[
\hat{U}_{\phi}\Psi=e^{-i\left(\phi/2\right)\sigma_{z}}R_{\phi}\Psi=\left[\begin{array}{c}
e^{-i\phi/2}\psi_{A}\left(r,\theta-\phi\right)\\
e^{i\phi/2}\psi_{B}\left(r,\theta-\phi\right)
\end{array}\right].
\]
Provided
\[
\left[\begin{array}{c}
\psi_{A}\left(r,\theta-\phi\right)\\
\psi_{B}\left(r,\theta-\phi\right)
\end{array}\right]=\left[\begin{array}{c}
e^{-im\theta}\psi_{A}\left(r,\theta\right)\\
e^{-i\left(m+1\right)\theta}\psi_{B}\left(r,\theta\right)
\end{array}\right]
\]
we get
\begin{equation}
\hat{U}_{\phi}\Psi=e^{-ik\phi}\Psi,
\end{equation}
with $k=m+1/2$. The only functions which are invariant under this
symmetry are linear combinations of $\left|z\right|^{2n}=r^{2n}$,
for $n\geq0$. For each $m$, the initial states that respect the
symmetries and the BC, $\psi_{B}/\psi_{A}=ie^{i\theta}$ are (with
$z=x+iy$ and $\bar{z}$ the complex conjugate of $z$)
\begin{align}
\Psi_{0}^{\left(m\ge0\right)}\left(z,\bar{z}\right) & =z^{m}\left[\begin{array}{c}
-i\\
z
\end{array}\right],\nonumber \\
\Psi_{0}^{\left(m<0\right)}\left(z,\bar{z}\right) & =\left(\bar{z}\right)^{-(m+1)}\left[\begin{array}{c}
\bar{z}\\
i
\end{array}\right].
\end{align}

With no loss of generality, we discuss the states with $m=0$. The
procedure is analogous for the remaining representations. A polynomial
state can have the form 
\begin{equation}
\ket{P\left(r\right),Q\left(r\right)}=\left[\begin{array}{c}
-iP\left(r\right)\\
zQ\left(r\right)
\end{array}\right]\label{eq:poly_state}
\end{equation}
if $P$ and $Q$ are even polynomials in $r$ such that $P\left(1\right)=Q\left(1\right)=1$,
thus respecting both the BC and the symmetries in question. Simple
monomials $P_{n}(r)=r^{2n}$ satisfy both conditions and a non-orthogonal
basis is 
\begin{equation}
\left\{ \ket{P_{0},P_{0}},\ket{P_{1},Q_{0}},\ket{P_{0},P_{1}},\ket{P_{2},P_{0}},\ket{P_{0},P_{2}},\dots\right\} \label{eq:poly_basis-1}
\end{equation}
 After applying the G-S process, the Hamiltonian is straightforwardly
computed and diagonalized for different basis sizes. 

The comparison with the exact results is presented in Table \ref{tab:Convergence-of-polynomial}.
As the method quickly converges towards the exact results, we will
now apply it to the square enclosure with the same BC.
\begin{table}[H]
\begin{centering}
\begin{tabular}{|c|c|c|c|c|}
\hline 
Basis Size & $\epsilon_{0}$ & $\epsilon_{1}$ & $\epsilon_{2}$ & $\epsilon_{3}$\tabularnewline
\hline 
\hline 
{\small 2} & {\footnotesize$1.4415$} & {\footnotesize$-2.77485$} &  & \tabularnewline
\hline 
{\small 3} & {\footnotesize$1.4344$} & {\footnotesize$-3.20749$} & {\footnotesize$4.17305$} & \tabularnewline
\hline 
{\small 5} & {\footnotesize$1.4347$} & {\footnotesize$-3.11371$} & {\footnotesize$4.61325$} & {\footnotesize$-7.05342$}\tabularnewline
\hline 
{\small 7} & {\footnotesize$1.4347$} & {\footnotesize$-3.11287$} & {\footnotesize$4.67908$} & {\footnotesize$-6.30658$}\tabularnewline
\hline 
Exact & {\footnotesize$1.4347$} & {\footnotesize$-3.11286$} & {\footnotesize$4.6801$} & {\footnotesize$-6.26629$}\tabularnewline
\hline 
\end{tabular}
\par\end{centering}
\caption{Convergence of polynomial method for the circular enclosure.\protect\label{tab:Convergence-of-polynomial}}
\end{table}

\subsection{Non-Dirichlet Square Enclosure}

The case of square with edges at $x,y=\pm L/2$ and $t=1$ has no
known analytic solution. As these boundary conditions correspond to
an uniform gap outside the enclosure, the Hamiltonian will be invariant
under $\pi/2$ rotations.
\[
\hat{U}=e^{-i\frac{\pi}{4}\sigma_{z}}R_{\pi/2}
\]
The one-dimensional representations of this symmetry are given by
\[
\hat{U}\Psi=e^{i\alpha}\Psi,
\]
where 
\begin{equation}
\alpha=\frac{\pi}{4}+n\frac{\pi}{2}\in\left[-\pi,\pi\right].
\end{equation}
The other possible symmetries of the Hamiltonian would be reflections.
These, however, change the sign of the mass gap outside the enclosure
and are also not compatible with the imposed boundary conditions. 

The following are low order polynomial functions, belonging to each
of the representations of the symmetry group.
\begin{align}
\Phi_{\pi/4} & =\left[\begin{array}{c}
\bar{z}\\
1
\end{array}\right], & \Phi_{-3\pi/4} & =\left[\begin{array}{c}
z\\
z^{2}
\end{array}\right],\nonumber \\
\Phi_{3\pi/4} & =\left[\begin{array}{c}
\bar{z}^{2}\\
\bar{z}
\end{array}\right], & \Phi_{-\pi/4} & =\left[\begin{array}{c}
1\\
z
\end{array}\right].\label{eq:states}
\end{align}
Each time we multiply one of these states by $z^{m}\bar{z}^{n}$,
$\alpha$ changes as 
\begin{equation}
\alpha\rightarrow\alpha+\left(n-m\right)\frac{\pi}{2}.\label{eq:invariant}
\end{equation}
We stay in the same irreducible representation, provided $\alpha$
is unchanged modulo $2\pi$,\emph{ i.e.}, if 
\begin{equation}
n-m=0\quad\mod4.\label{eq:inv_cond}
\end{equation}
Invariant monomials of $z$ and $\bar{z}$ are $\left|z\right|^{2},z^{4},\left(\bar{z}\right)^{4}$
and their products.  There are two issues that need to be addressed
at this point: 
\begin{description}
\item [{(a)}] the states of Eq.(\ref{eq:states}) do not satisfy the BC,
and are therefore not suitable as starting states upon which to build
a basis. Taking the $\alpha=-\pi/4$ as an example, we require states
of the form 
\begin{equation}
\Phi_{0}=\left[\begin{array}{c}
P_{0}\left(\left|z\right|^{2},z^{4},\left(\bar{z}\right)^{4}\right)\\
zQ_{0}\left(\left|z\right|^{2},z^{4},\left(\bar{z}\right)^{4}\right)
\end{array}\right]
\end{equation}
where the invariant polynomials, satisfy $P_{0}-Q_{0}z=1$ for $y=-1/2$.
If this is verified at this edge, the symmetry will ensure the BC
are also verified at the other edges. The choice
\begin{align}
P_{0} & =-\frac{i}{5}-iz^{4}+\frac{i}{5}\bar{z}^{4}, & Q_{0} & =1+\frac{z^{4}}{5}-2\left|z\right|^{2}-\frac{1}{5}\bar{z}^{4}
\end{align}
 proved convenient in the sense that the behaviour of the ket $\Phi_{0}(x,y)$
near the origin resembles that of the lowest eigenvalue of the circular
enclosure,
\item [{(b)}] the basis must be built by multiplying each component of
the spinor by increasing order invariant polynomials which are required
to be unity at the edges of the enclosure so that the BC are preserved.
Such polynomials can be generated order by order, by imposing $P_{n}\left(\left|z\right|^{2},z^{4},\left(\bar{z}\right)^{4}\right)=1$
for $y=-1/2$.
\end{description}
Once these steps are performed, the calculation proceeds as for the
circle, by orthogonalizing the basis and calculating of the Hamiltonian
matrix, separately for each of the 4 symmetry representations. The
necessary work is halved because the transformation $\sigma_{x}\mathcal{K}$
($\mathcal{K}$ is the complex conjugation operator) changes the sign
of the Hamiltonian and preserves the BC; the states of the representations
with $\alpha=1/4,3\pi/4$ can be obtained from those of $\alpha=-1/4,-3\pi/4$
by this transformation; the eigenvalue spectrum remains symmetrical.

\begin{figure}[H]
\centering{}\includegraphics[scale=0.3]{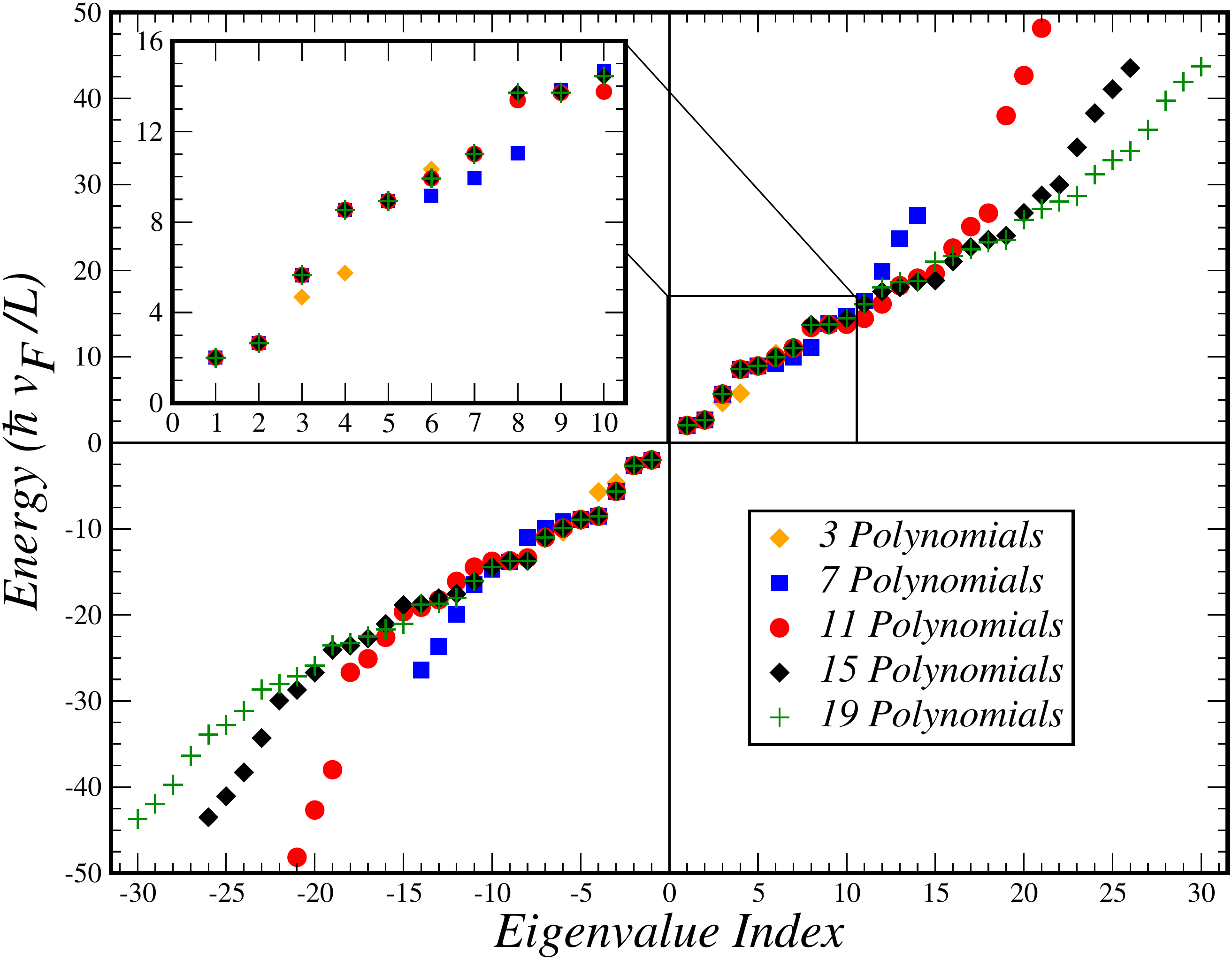}\caption{Comparison of the spectrum of the uniform-BCs square billiard with
increasing number of polynomials.\protect\label{fig:Spectrum-of-theuniformBCsquare}}
\end{figure}

The spectrum is displayed in Fig.\ref{fig:Spectrum-of-theuniformBCsquare},
with the low-energy regime in the inset. The number of polynomials
in the legend refers to the basis size in each irreducible representation.
 This method again converges, as it did in case of the circular enclosure. 

\begin{figure}[H]
\begin{centering}
\includegraphics[height=3.5cm]{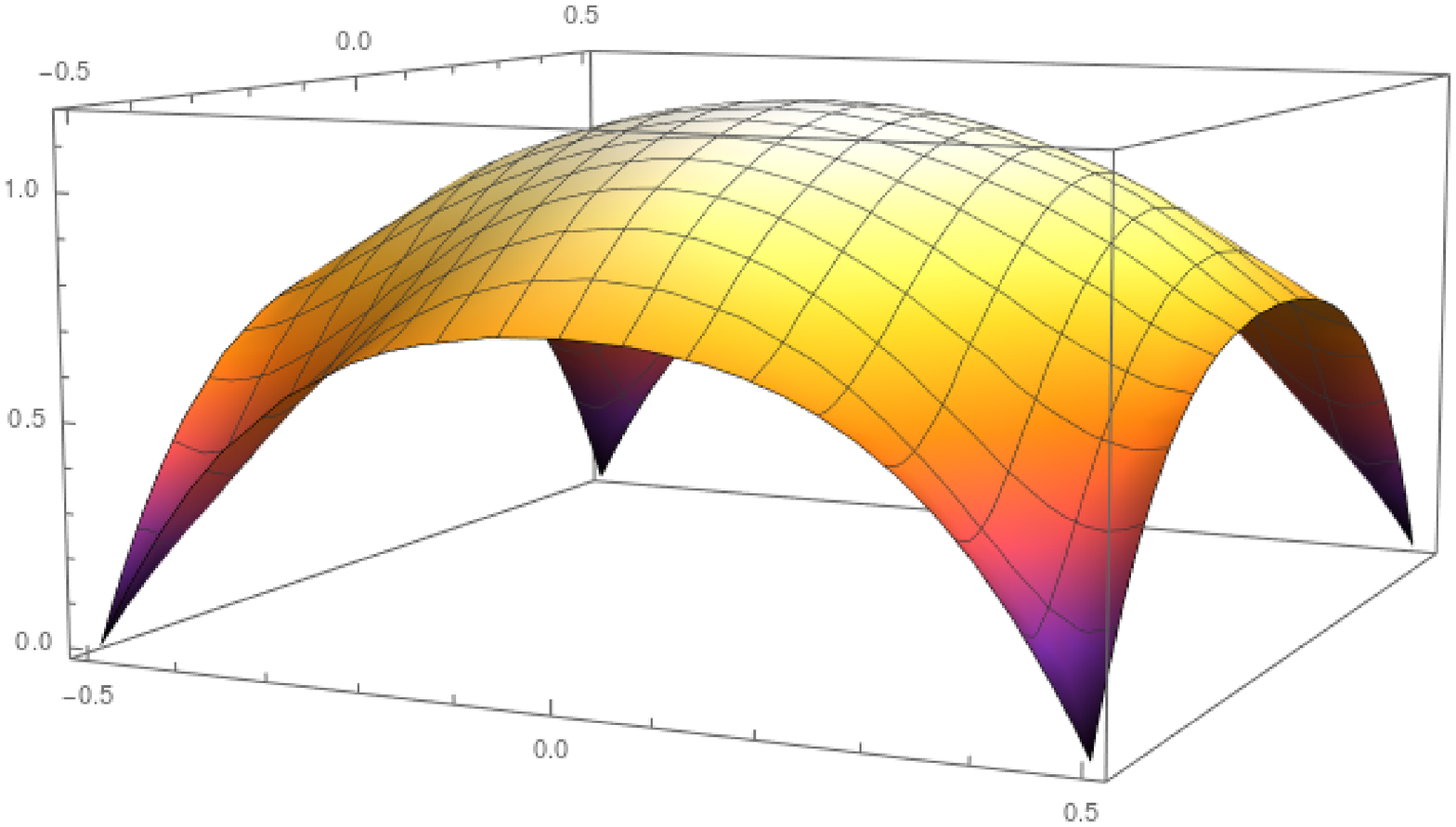}\includegraphics[height=3.5cm]{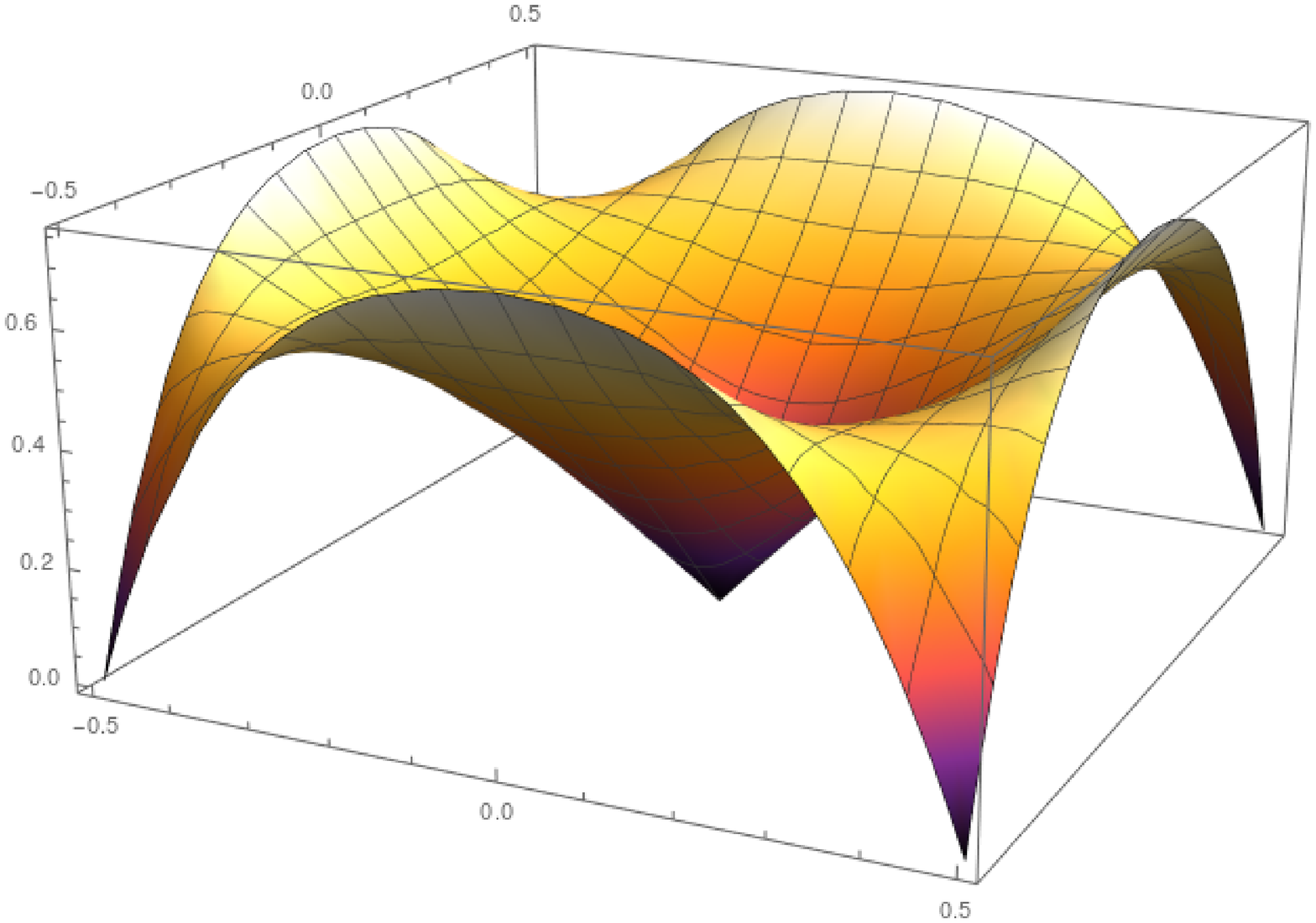}\includegraphics[height=3.5cm]{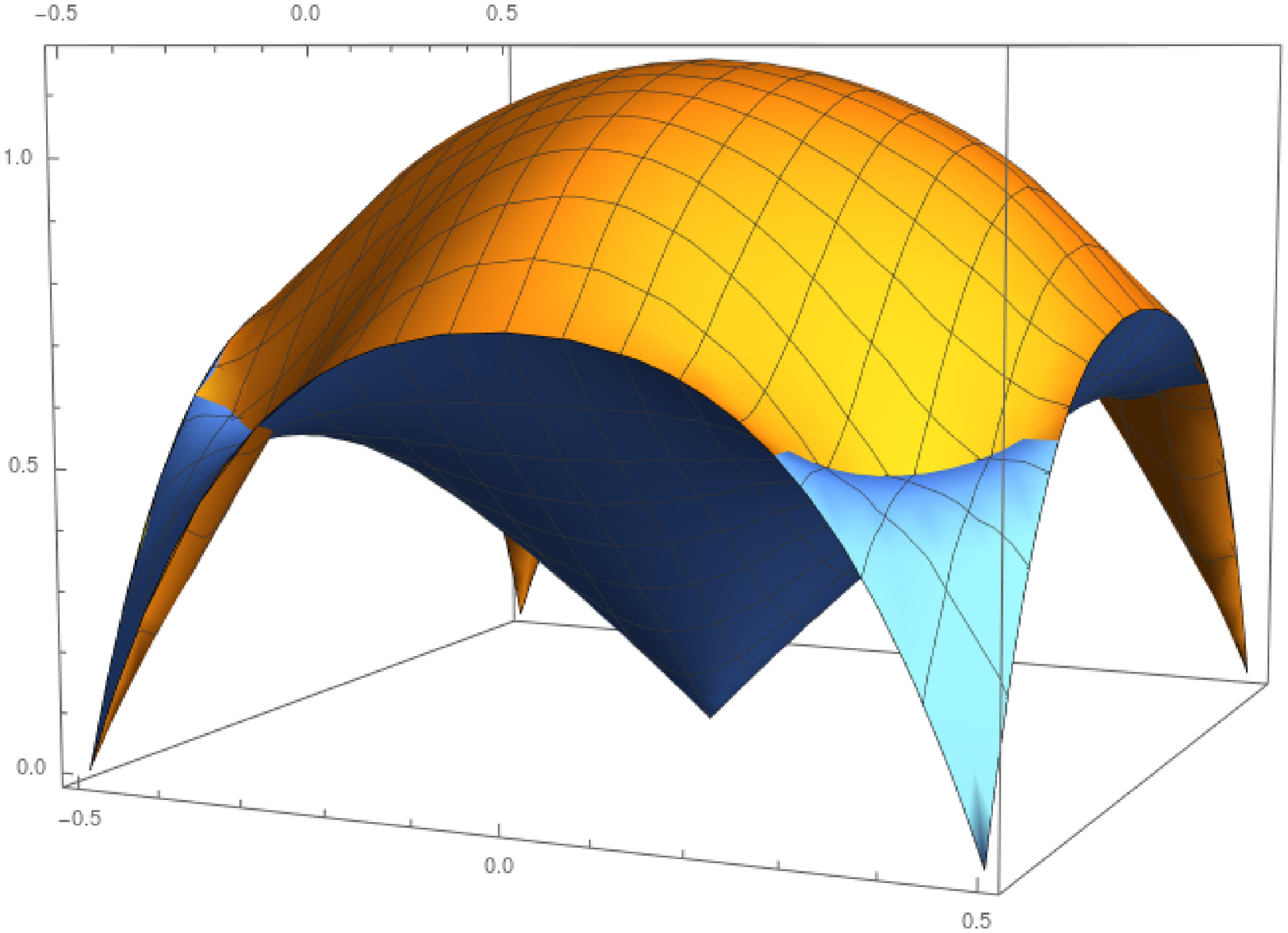}
\par\end{centering}
\caption{\protect\label{fig:Absolute-values-of}Absolute values of the spinor
wave-functions for the eigenstate of energy $\epsilon=-2.0\hbar v_{F}/L$.
Left panel, $\left|\psi_{A}\right|;$ middle,$\left|\psi_{B}\right|$;
right, both components, showing that $\left|\psi_{A}\right|=\left|\psi_{B}\right|$
at the edges as required by the BC. Near the center of the square
the behaviour resembles that of the corresponding state for the circle,
but here the spinor wavefunction is zero at the vertexes, as this
is the only way to satisfy the BC for the two edges that meet at the
vertex (colour online). }
\end{figure}

As stated before, the method also provides the eigenstates as polynomials.
The valence band state with energy closest to zero is plotted in Fig.\ref{fig:Absolute-values-of}.
At the center of the square, the behaviour of the two spinor components
is not unlike that of the corresponding state of the circle, but is
quite different at the edges. The two components have the same absolute
value at the edges, as required by the BC, and both vanish at the
corners, as that is the only way the BC of the edges that meet at
the corner can be satisfied. 

To finalize the analysis of the method, we tested it in a problem
with BCs analogous to those found in zigzag-terminated hexagonal graphene
flakes. 

\subsection{Dirichlet Hexagonal Enclosure \protect\label{subsec:Dirichlet-Hexagonal-Enclosure}}

We applied this method to the study of an hexagonal enclosure with
the BCs equivalent to those of an hexagonal graphene flake whose edges
are all zigzag-terminated. This specific set of BCs can be expressed
by alternating $t$ between $t=0$ and $t\rightarrow\infty$, depending
on the termination sublattice.

As we are working with Dirichlet boundaries in alternating sides for
each spinor component, we can construct the initial polynomial as
in equation \ref{eq:square_poly}. Considering a regular hexagon centered
at the origin of side-length $L$, the two spinor components for the
starting state will be defined as 
\begin{align}
\psi_{0,A}\left(x,y\right) & =\left[\frac{\sqrt{3}L}{2}+y\right]\left[L+\left(x-\frac{y}{\sqrt{3}}\right)\right]\nonumber \\
 & \ \ \ \times\left[L-\left(x+\frac{y}{\sqrt{3}}\right)\right]\label{eq:hex_poly}\\
\psi_{0,B}\left(x,y\right) & =\psi_{A}\left(x,-y\right).\nonumber 
\end{align}

With this, we diagonalize the Hamiltonian 
\begin{align}
\bra{\Psi_{i}}H^{\dagger}H\ket{\Psi_{j}} & =\bra{\psi_{i,A}}\left(-\hbar^{2}v_{F}^{2}\nabla^{2}\right)\ket{\psi_{j,A}}+\nonumber \\
 & \ \ \ +\bra{\psi_{i,B}}\left(-\hbar^{2}v_{F}^{2}\nabla^{2}\right)\ket{\psi_{j,B}}.\label{eq:hamilt}
\end{align}
\begin{figure}[H]
\centering{}\includegraphics[scale=0.33]{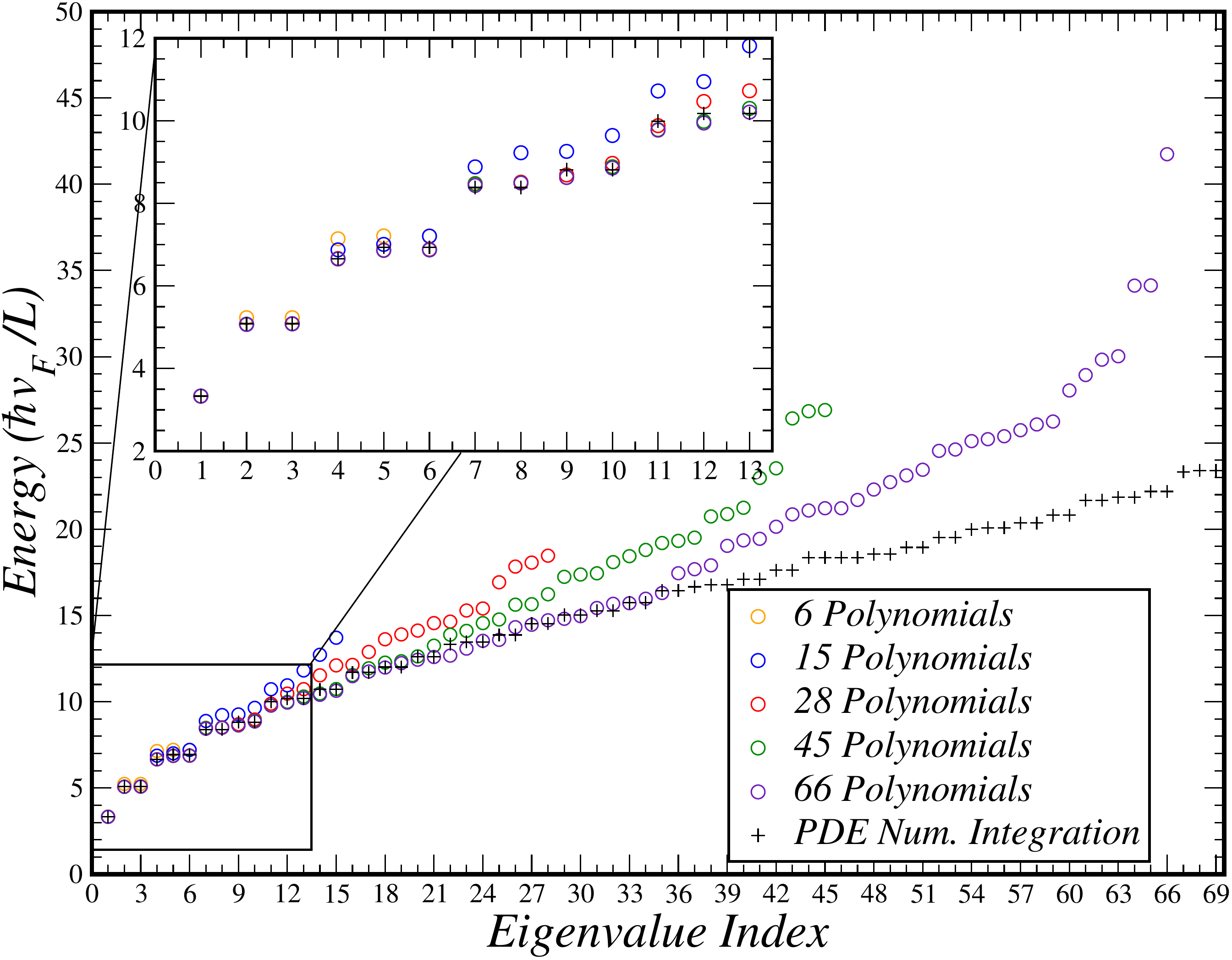}\caption{Comparison of the spectrum of the zigzag-like hexagonal enclosure
with increasing number of polynomials.\protect\label{fig:Spectrum-of-the-zigzag-hex-helmholtz-1}}
\end{figure}
The resulting spectrum is visible in Fig.\ref{fig:Spectrum-of-the-zigzag-hex-helmholtz-1}
for different basis sizes. The numerical solution was obtained via
\textit{Wolfram Mathematica}\textsuperscript{\textregistered} by
imposing Dirichlet BCs on three alternating sides of the enclosure
in question, leaving the remaining free. 

\begin{figure}[H]
\begin{spacing}{0.7}
\noindent \raggedright{}%
\begin{minipage}[t]{0.3\textwidth}%
\begin{center}
\includegraphics[scale=0.2]{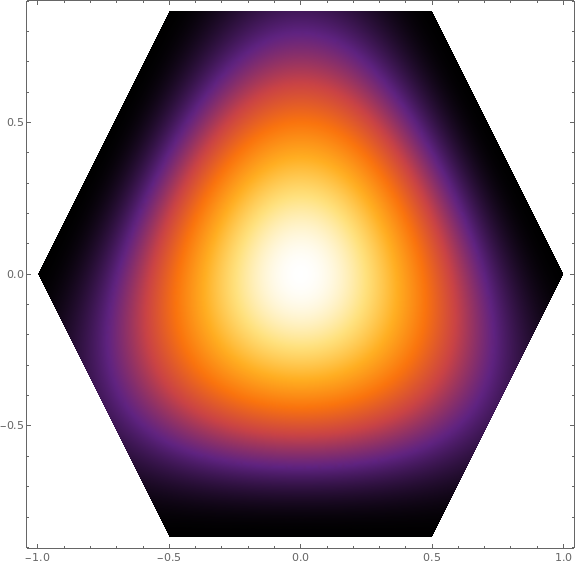}
\par\end{center}%
\end{minipage}\hfill{}%
\begin{minipage}[t]{0.3\textwidth}%
\begin{center}
\includegraphics[scale=0.2]{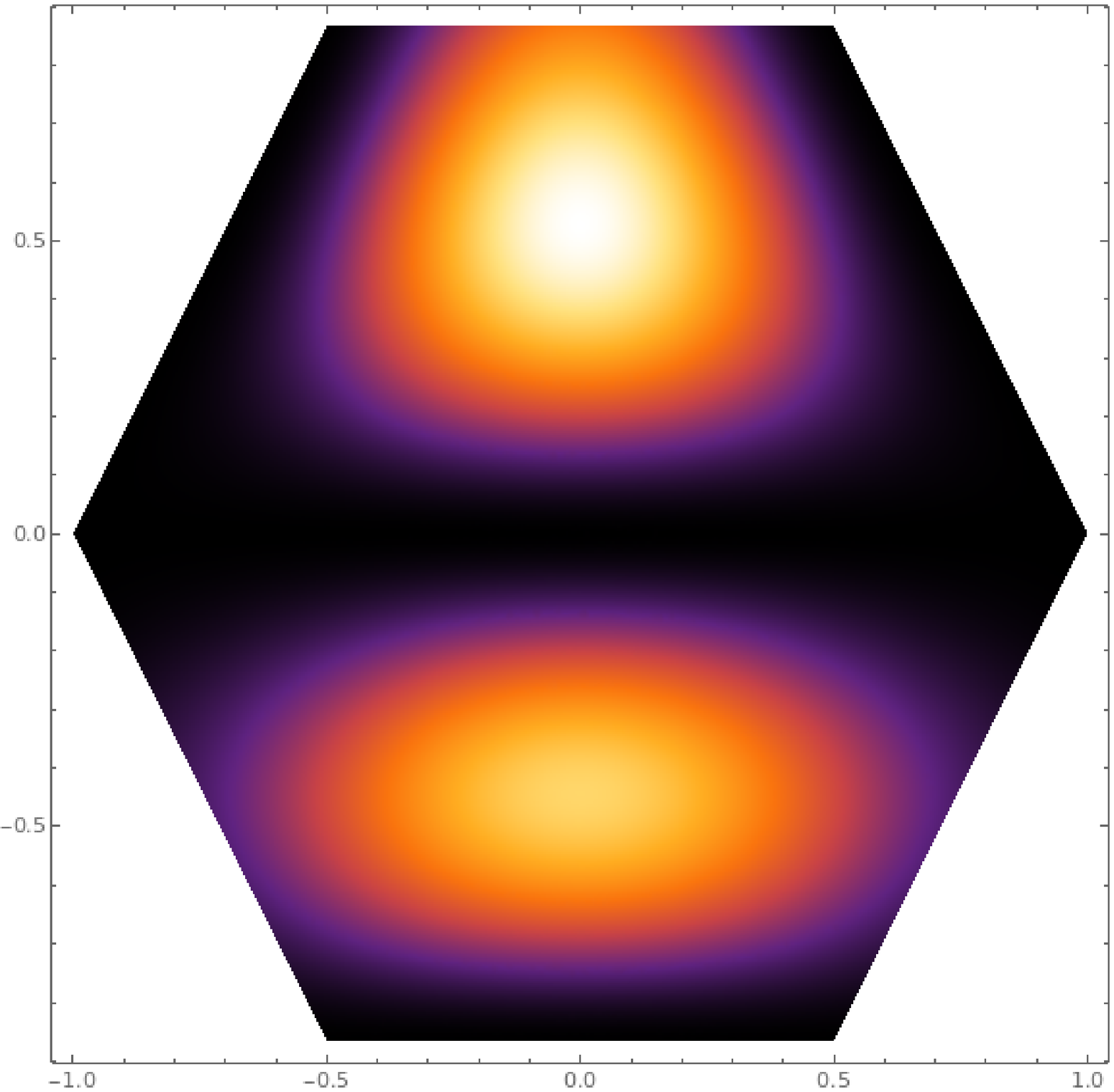}
\par\end{center}%
\end{minipage}\hfill{}%
\begin{minipage}[t]{0.3\textwidth}%
\begin{center}
\includegraphics[scale=0.2]{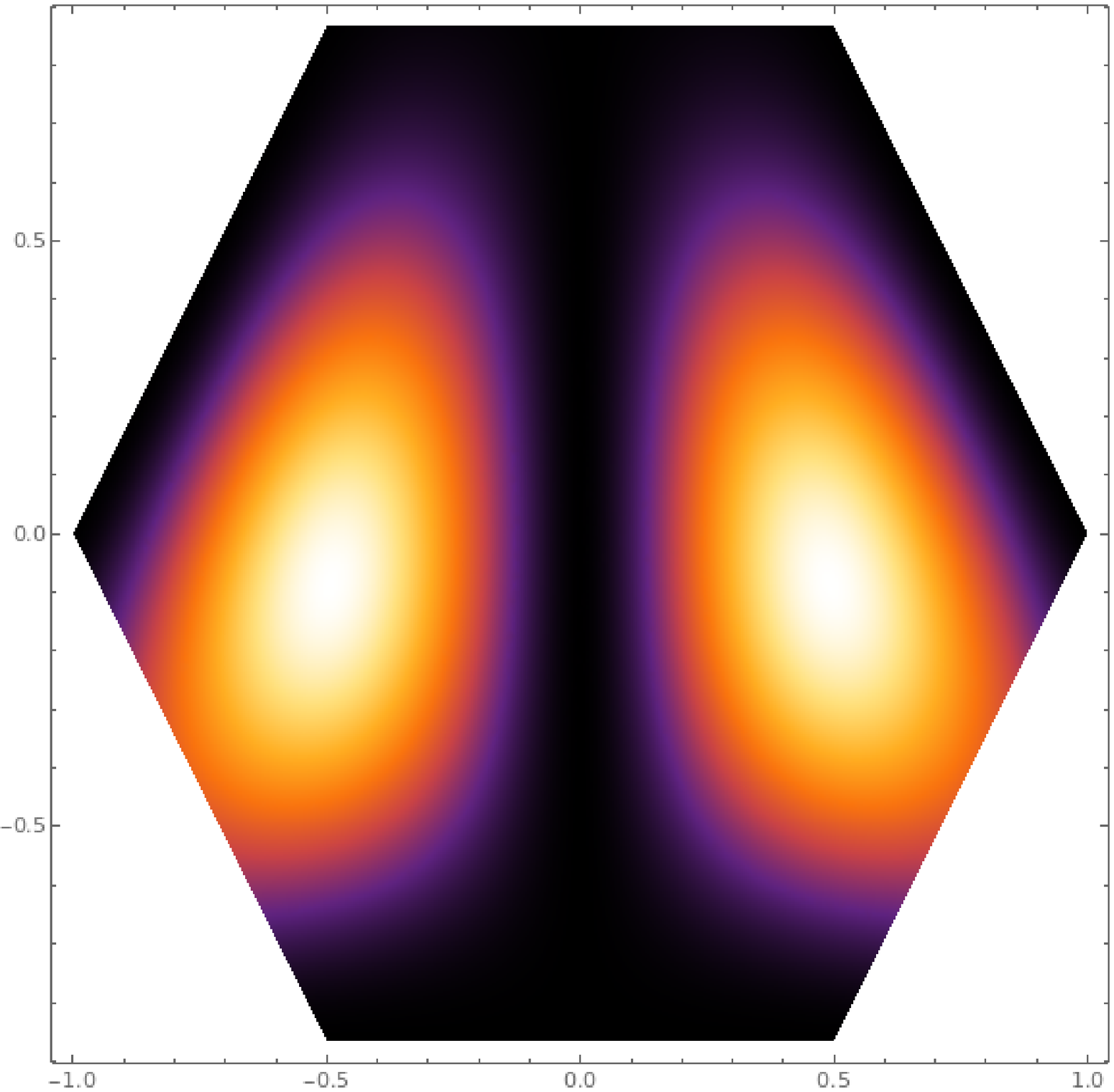}
\par\end{center}%
\end{minipage}\hfill{}%
\begin{minipage}[t]{0.3\textwidth}%
\begin{center}
\includegraphics[scale=0.2]{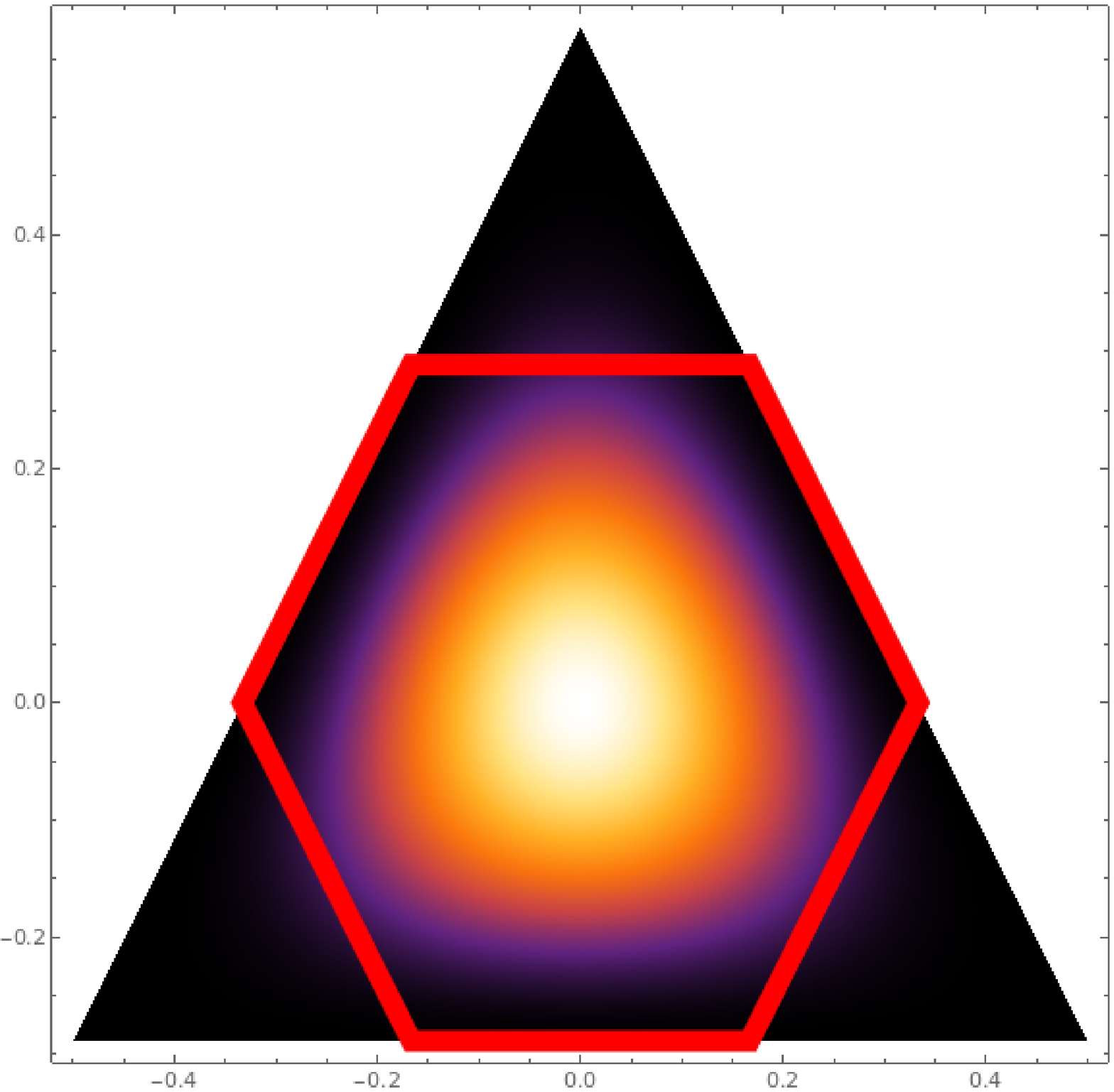}
\par\end{center}%
\end{minipage}\hfill{}%
\begin{minipage}[t]{0.3\textwidth}%
\begin{center}
\includegraphics[scale=0.2]{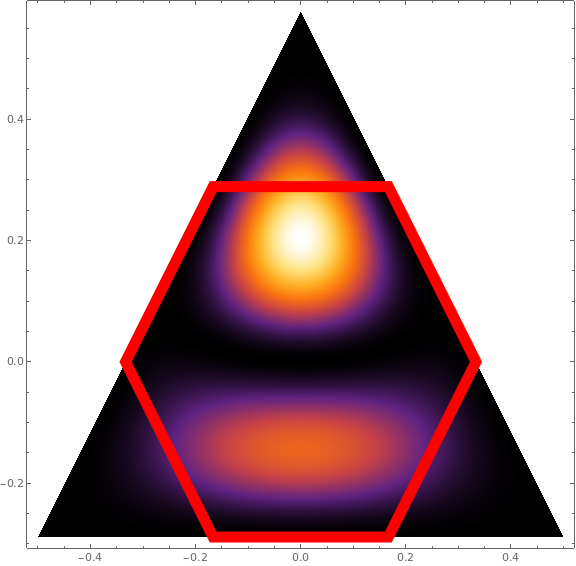}
\par\end{center}%
\end{minipage}\hfill{}%
\begin{minipage}[t]{0.3\textwidth}%
\begin{center}
\includegraphics[scale=0.2]{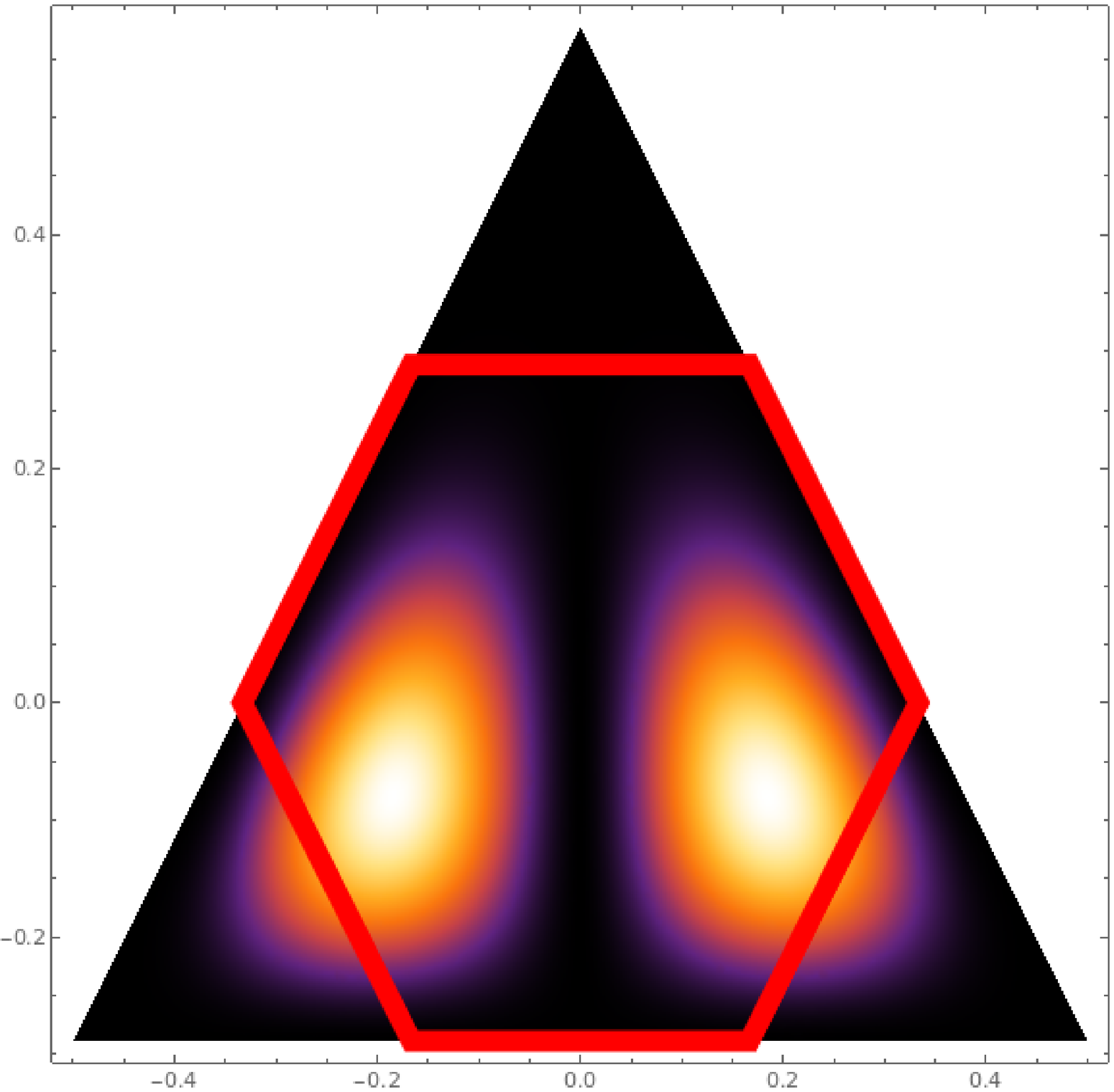}
\par\end{center}%
\end{minipage}\\
\begin{minipage}[c]{0.5\textwidth}%
\noindent \begin{flushleft}
\includegraphics[scale=0.6]{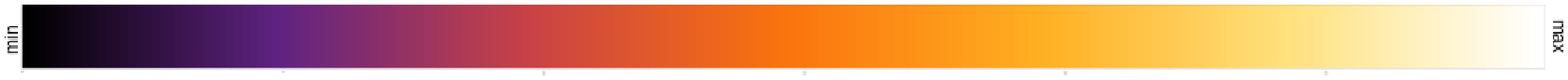}
\par\end{flushleft}%
\end{minipage}\caption{$\left|\psi_{A,j}\right|^{2}$ for the hexagon (top) and for the triangle
(bottom) $j\in\left\{ 1,2,3\right\} $ (color online).\protect\label{fig:-for-the_hex-vs-triang}}
\end{spacing}
\end{figure}

\begin{figure}[H]
\centering{}\includegraphics[scale=0.33]{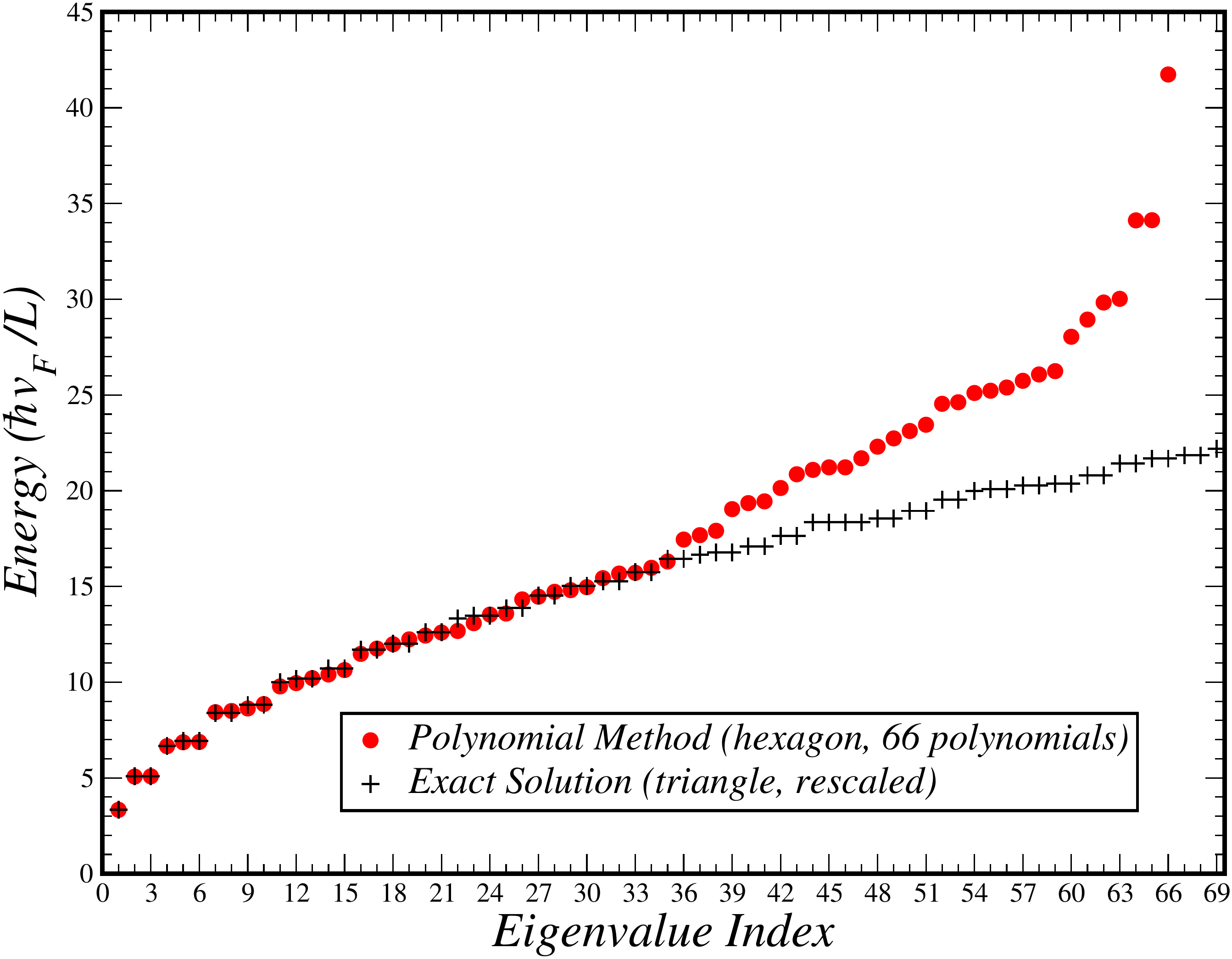}\caption{Comparison of the spectrum of the zigzag-like hexagonal billiard against
the exact solution for the triangle.\protect\label{fig:Spectrum-of-the-zigzag-hex-helmholtz-vs-triang}}
\end{figure}

Finally, we will compare these results with the ones obtained by Gaddah
{[}\citealp{gaddah_exact_2018}{]} for the problem of the triangular
billiards with boundary condition $\psi_{A}=0$ (in our notation,
$t\rightarrow\infty$). The exact spectrum described by the author
follows the relation ($L$ is the side-length of the equilateral triangle
in question)
\begin{align}
E_{n_{1},n_{2}} & =\pm\frac{4}{3}\frac{\pi}{L}\hbar v_{F}\sqrt{n_{1}^{2}+n_{1}n_{2}+n_{2}^{2}},
\end{align}
where $n_{2}\geq n_{1}>0$. As described in the article, the states
with $n_{2}>n_{1}$ are (at least) $2\times$ degenerate. The BCs
we imposed on a hexagon, $\psi_{A}=0$ on three alternating sides
(Eq.(\ref{eq:hex_poly})), also enforce $\psi_{A}=0$ on the edges
of an equilateral triangle, that contains the hexagon region in question.
The condition for $\psi_{B}$ is the same, as it is a simple inversion
of this same triangle. As a consequence, after re-scaling the exact
spectrum for the triangle by $\sqrt{A_{\mathrm{hex}}/A_{\mathrm{triang}}}$,
we obtain a match between the two as shown in Fig.\ref{fig:Spectrum-of-the-zigzag-hex-helmholtz-vs-triang}.
We also compare the density plots for the first three approximate
eigenfunctions to those present in Gaddah's work. This comparison
(using 10 polynomials)  is presented in Fig.\ref{fig:-for-the_hex-vs-triang}
for $\left|\psi_{A}\right|^{2}$, with the corresponding regions highlighted.


\section{Summary and Conclusions}

With the polynomial method we have been able to construct approximate
solutions to Schr{\"o}dinger and Dirac-Weyl equations for planar
convex polygonal enclosures. The method has been applied before to
the determination of frequencies of plate vibrations in the Mechanical
Engineering community, but as far as we know its generalization to
the Dirac billiards has not been reported before. In the this case
we considered different types of boundary conditions, including situations
where only the ratio of spinors components is specified. 

The method is based on constructing a truncated basis of states, all
of which satisfy the prescribed BC from the start. By comparing with
cases where the exact solution is known (most often by separation
of variables) we were able to ascertain the convergence of the method.
Typically, a basis size of order $3n$ is required to obtain $n$
eigenvalues with close to 1\% accuracy. The method provides not only
the lowest energies (or lowest absolute energies for the Dirac -Weyl
case) but also eigenstates as finite order polynomials in the coordinates. 

\section*{Acknowledgments}

\begin{singlespace}
The authors acknowledge financing of Fundação da Ciência e Tecnologia,
of COMPETE 2020 program in FEDER component (European Union), through
projects POCI-01-0145-FEDER-028887 and UID/FIS/04650/2019. The authors
also acknowledge financial support from Fundação para a Ciência e
Tecnologia, Portugal, through national funds, co-financed by COMPETE-FEDER
(grant M-ERA-NET2/0002/2016 -- UltraGraf) under the Partnership Agreement
PT2020.
\end{singlespace}

\bibliographystyle{unsrt}
\phantomsection\addcontentsline{toc}{section}{\refname}\bibliography{article_jls_V9_1col.bbl}

\begin{thebibliography}{10}

\bibitem{Lame1852}
G.~Lam{\'e}.
\newblock {Leçons sur la Théorie Mathématique de l'Elasticité}.
\newblock {\em {Bachelier}, {1852}}.

\bibitem{ISI:A1980KL96700005}
M.A. Pinsky.
\newblock {The Eigenvalues of An Equilateral Triangle}.
\newblock {\em {Siam Journal On Mathematical Analysis}}, {11}({5}):{819--827},
  {1980}.

\bibitem{ISI:000183800600005}
BJ~McCartin.
\newblock {Eigenstructure of the equilateral triangle, Part I: The Dirichlet
  problem}.
\newblock {\em {Siam Review}}, {45}({2}):{267--287}, {JUN} {2003}.

\bibitem{ISI:A1991GC36900022}
V~Amar, M~Pauri, and A~Scotti.
\newblock {{Schr{\"o}dinger}-equation For Convex Plane Polygons - A Tiling
  Method For The Derivation Of Eigenvalues And Eigenfunctions}.
\newblock {\em {Journal of Mathematical Physics}}, {32}({9}):{2442--2449},
  {Sep} {1991}.

\bibitem{li_particle_1987}
Wai-Kee Li and S.~M. Blinder.
\newblock Particle in an equilateral triangle: {Exact} solution of a
  nonseparable problem.
\newblock {\em Journal of Chemical Education}, 64(2):130, February 1987.

\bibitem{gaddah_lie_2013}
W.~A. Gaddah.
\newblock A {Lie} group approach to the {Schr{\"o}dinger} equation for a
  particle in an equilateral triangular infinite well.
\newblock {\em European Journal of Physics}, 34(5):1175--1186, July 2013.

\bibitem{ISI:A1993LQ34500002}
V~Amar, M~Pauri, and A~Scotti.
\newblock {{Schr{\"o}dinger}-equation For Convex Plane Polygons .2. A No-go
  Theorem For Plane-waves Representation Of Solutions}.
\newblock {\em {Journal Of Mathematical Physics}}, {34}({8}):{3343--3350},
  {Aug} {1993}.

\bibitem{berry_michael_victor_neutrino_1987}
{Berry Michael Victor} and {Mondragon R. J.}
\newblock Neutrino billiards: time-reversal symmetry-breaking without magnetic
  fields.
\newblock {\em Proceedings of the Royal Society of London. A. Mathematical and
  Physical Sciences}, 412(1842):53--74, July 1987.

\bibitem{novoselov_two-dimensional_2005}
K.~S. Novoselov, A.~K. Geim, S.~V. Morozov, D.~Jiang, M.~I. Katsnelson, I.~V.
  Grigorieva, S.~V. Dubonos, and A.~A. Firsov.
\newblock Two-dimensional gas of massless {Dirac} fermions in graphene.
\newblock {\em Nature}, 438(7065):197--200, November 2005.

\bibitem{castro_neto_electronic_2009}
A.~H. Castro~Neto, F.~Guinea, N.~M.~R. Peres, K.~S. Novoselov, and A.~K. Geim.
\newblock The electronic properties of graphene.
\newblock {\em Reviews of Modern Physics}, 81(1):109--162, January 2009.

\bibitem{ponomarenko_chaotic_2008}
L.~A. Ponomarenko, F.~Schedin, M.~I. Katsnelson, R.~Yang, E.~W. Hill, K.~S.
  Novoselov, and A.~K. Geim.
\newblock Chaotic {Dirac} {Billiard} in {Graphene} {Quantum} {Dots}.
\newblock {\em Science}, 320(5874):356--358, April 2008.

\bibitem{libisch_graphene_2009}
Florian Libisch, Christoph Stampfer, and Joachim Burgdörfer.
\newblock Graphene quantum dots: {Beyond} a {Dirac} billiard.
\newblock {\em Physical Review B}, 79(11):115423, March 2009.

\bibitem{zarenia_energy_2011}
M.~Zarenia, A.~Chaves, G.~A. Farias, and F.~M. Peeters.
\newblock Energy levels of triangular and hexagonal graphene quantum dots: a
  comparative study between the tight-binding and the {Dirac} approach.
\newblock {\em Physical Review B}, 84(24):245403, December 2011.

\bibitem{gaddah_exact_2018}
W.~A. Gaddah.
\newblock Exact solutions to the {Dirac} equation for equilateral triangular
  billiard systems.
\newblock {\em Journal of Physics A: Mathematical and Theoretical},
  51(38):385304, September 2018.

\bibitem{ISI:000289149100002}
Liang Huang, Ying-Cheng Lai, and Celso Grebogi.
\newblock {Characteristics of level-spacing statistics in chaotic graphene
  billiards}.
\newblock {\em {CHAOS}}, {21}({1}), {MAR} {2011}.

\bibitem{ISI:000521256300001}
Liang Huang and Ying-Cheng Lai.
\newblock {Perspectives on relativistic quantum chaos}.
\newblock {\em {Communications in Theoretical Physics}}, {72}({4}), {APR 1}
  {2020}.

\bibitem{ISI:A1995RG85100019}
Y~Shimizu and A~Shudo.
\newblock {Polygonal Billiards - Correspondence Between Classical Trajectories
  And Quantum Eigenstates}.
\newblock {\em {Chaos Solitons \& Fractals}}, {5}({7}):{1337--1362}, {Jul}
  {1995}.

\bibitem{bhat_flexural_1987}
R.B. Bhat.
\newblock Flexural vibration of polygonal plates using characteristic
  orthogonal polynomials in two variables.
\newblock {\em Journal of Sound and Vibration}, 114(1):65--71, January 1987.

\bibitem{liew_free_1990}
K.~M. Liew, K.~Y. Lam, and S.~T. Chow.
\newblock Free vibration analysis of rectangular plates using orthogonal plate
  function.
\newblock {\em Computers \& Structures}, 34(1):79--85, January 1990.

\bibitem{liew_set_1991}
K.~M. Liew and K.~Y. Lam.
\newblock A {Set} of {Orthogonal} {Plate} {Functions} for {Flexural}
  {Vibration} of {Regular} {Polygonal} {Plates}.
\newblock {\em Journal of Vibration and Acoustics}, 113(2):182--186, April
  1991.

\bibitem{larcher_notes_1959}
H.~Larcher.
\newblock Notes on orthogonal polynomials in two variables.
\newblock {\em Proceedings of the American Mathematical Society},
  10(3):417--423, 1959.

\bibitem{Quintela2019}
M.~Quintela.
\newblock From the {1D} {Schr{\"o}dinger} infinite well to {Dirac-Weyl}
  graphene flakes.
\newblock Master's thesis, Faculdade de Ciências, Universidade do Porto, 2019.

\bibitem{PhysRevB.77.035316}
B.~Wunsch, T.~Stauber, and F.~Guinea.
\newblock Electron-electron interactions and charging effects in graphene
  quantum dots.
\newblock {\em Phys. Rev. B}, 77:035316, Jan 2008.

\end{thebibliography}

\end{document}